\documentclass[12pt]{article}

\usepackage{latexsym,amsmath,amssymb,theorem,epsfig}

\usepackage{cite}

\DeclareFontFamily{OT1}{rsfs10}{}
\DeclareFontShape{OT1}{rsfs10}{m}{n}{ <-> rsfs10 }{}
\DeclareMathAlphabet{\mathscript}{OT1}{rsfs10}{m}{n}

\numberwithin{equation}{section}

\newcommand{\ns}{\normalsize}

\newcommand{\tr}{\text{tr}}
\newcommand{\pt}{\partial}

\def\a{\alpha}
\def\b{\beta}
\def\g{\gamma}

\def\d{\delta}

\def\k{\kappa}
\def\l{\lambda}

\def\s{\sigma}
\def\t{\tau}

\def\D{\Delta}

\def\O{\Omega}

\def\gsim{ \lower .75ex \hbox{$\sim$} \llap{\raise .27ex \hbox{$>$}} }
\def\lsim{ \lower .75ex \hbox{$\sim$} \llap{\raise .27ex \hbox{$<$}} }
\def\be{\begin{equation}}
\def\ee{\end{equation}}
\def\bea{\begin{eqnarray}}
\def\eea{\end{eqnarray}}

\def \td {\tilde}

\def \V  {{\rm V}}

\def \sql {{\sqrt{\l}}\ }
\def \del{\partial}
\def \a {\alpha}

\def\ov{\over}
\def \ci {\cite}

\def \foot {\footnote}
\def \bi{\bibitem}
\def\la{\label}\def\foot{\footnote}\newcommand{\rf}[1]{(\ref{#1})}

\def \no {\nonumber}

\def \adss {$AdS_5 \times S^5\ $}

\def \a {\alpha } 
\def \eps {\epsilon}

\def \GG  {{\cal G}}
\def \V  {{\rm V}}\def \UU {{v}}

\def\ket {\big\rangle}
\def\bra {\big\langle }
\def \VV {{\cal V}}

\def \oc {\Omega_c} 
\def \tX {\tilde X}
\def \tvarphi {\tilde \varphi}
\def \tpsi {\tilde \psi}

\def \nN {{\rm N}}

\theoremstyle{plain}

{\theorembodyfont{\rmfamily} }

\def \ed {\end{document}}

\begin{document}

\begin{titlepage}

\vspace{-5cm}


\vspace{-5cm}

\title{
  \hfill{\small Imperial-TP-EIB-2011-02  }  \\[1em]
   {\LARGE   Semiclassical correlators of three  states 
    with   large\\ $S^5$ charges 
    in string theory in $AdS_5 \times S^5$
}
\\[1em] }
\author{
   E.I. Buchbinder\footnote{e.buchbinder@imperial.ac.uk}  and 
     A.A. Tseytlin\footnote{Also at Lebedev  Institute, Moscow. 
   tseytlin@imperial.ac.uk }
     \\[0.5em]
   {\ns The Blackett Laboratory, Imperial College, London SW7 2AZ, U.K.}} 


\date{}

\maketitle

\begin{abstract}
We consider semiclassical  computation of 3-point  correlation functions
of (BPS or non-BPS) string states represented by vertex operators 
carrying large charges in $S^5$. 
We argue that the $AdS_5$  part  of the  construction 
of relevant semiclassical  solution  involves two basic  ingredients:
(i) configuration of three glued geodesics in $AdS_2$ 
  suggested by Klose and McLoughlin in arXiv:1106.0495
and (ii) a particular Schwarz-Christoffel map 
of  the 3-geodesic solution in  cylindrical 
($\tau,\sigma)$ domain into the  complex plane with three marked points.
This map is  constructed using the expression for the $AdS_2$ string stress tensor 
which is uniquely determined by  the 3 scaling dimensions $\D_i$
as noted  by  Janik and Wereszczynski in arXiv:1109.6262
(our solution, however, is different from theirs). 
We also  find  the $S^5$ part of the solution  and thus the full 
expression for the 
semiclassical part of the 3-point correlator for several examples:  
extremal and non-extremal  correlators  of BPS states  and 
a particular correlator of  ``small'' circular spinning strings in $S^3 \subset S^5$. 
We demonstrate  that for  the BPS  correlators  the results agree 
with the  large charge limit  of the corresponding 
supergravity and free gauge theory expressions.


\end{abstract}

\thispagestyle{empty}

\end{titlepage}

\section{Introduction}


One of the  central  problems  in  solving  conformal planar 
 ${\cal N}=4$ super Yang-Mills theory being guided by
gauge-string duality is to compute 3-point correlation  functions of conformal
primary operators at any value   of gauge coupling. Recently, some progress was
achieved in understanding  correlators of certain operators with
large quantum numbers at strong coupling  using semiclassical string theory 
approach (see, e.g.,  \ci{j,bt1,z,v,km,ja}  and references there).
The form of a 3-point function of  scalar primary operators is fixed by conformal
invariance to be 
$G=\  C\ |\vec a_{12}|^{-\a_3}|\vec a_{23}|^{-\a_1}|\vec a_{31}|^{-\a_2}  $ where $\vec a_{ij}=\vec 
a_i-\vec
a_j$ are  differences of 4-coordinates  and 
$\a_1= \D_2+\D_3 - \D_1$, etc.   are determined   by the three conformal dimensions $\D_i$. The  
 coefficient $C$ is a function of $\D_i$ and other quantum
numbers of the operators  and  also   depends on  `t Hooft coupling $\l$ or  string
tension $\sql \ov 2 \pi$. 
On the string theory side 
$G $   is defined as a correlator of the corresponding  vertex operators. 
For  $\l \gg 1$ and   when all three sets of quantum   numbers 
are semiclassically large (i.e. of order $\sql$)
 one may expect  $C$  to be given by a semiclassical 
approximation to the string path integral, thus  scaling as $e^{-\sql A}$ where 
$A$  is a function of the semiclassical parameters $d_i = { \D_i \ov \sql}, $ etc. 
 The 
semiclassical trajectory  should solve the string equations with  ``sources'' 
prescribed by the vertex operators. Finding such a  solution in general appears to be
 non-trivial. 

It is  natural to start with 
 a correlator of three   1/2 BPS operators  with large charges and
  dimensions, 
$\D_i = J_i \gg 1$.\foot{Below we shall only  consider  the leading order in large charge limit 
and thus will ignore  possible  difference   between 
the  dimensions $\D_i$ and the charges $J_i$.}
In this case the 3-point  correlator does not non-trivially 
depend on $\l$,  with  $C$ being a particular function of 
the quantum numbers only \ci{Seiberg}.
One may then  try to reproduce the expected large charge 
limit  of $C$  using semiclassical string theory
arguments. As the 
semiclassical limit of  the 
2-point correlator of BPS operators  is determined by a
 euclidean 
continuation of  a massless geodesic in \adss 
\ci{pol,t,yo,j} 
 one may expect that  in this case the relevant 
semiclassical trajectory should   be given 
by an intersection of the three 
 geodesics  \ci{km}
(with an intersection point being in the bulk of 
$AdS_5$ in the non-extremal case of $\D_1 \not= \D_2 + \D_3$.)

At the same time, the non-renormalization of the 3-point function of the BPS operators 
implies that it is given simply
by the supergravity expression 
and thus its large charge 
asymptotics can be captured 
 \ci{bt2} just by a  stationary point approximation of the 
supergravity
integral of the product of the 
corresponding  wave functions   over $AdS_5 \times S^5$.
This integral may be viewed as a localization
 of the string path integral 
where the string is shrunk 
to a point and one integrates over the 0-mode (center-of-mass point) 
only.

Below  we will use this  supergravity picture as a guide to arrive 
at a consistent semiclassical string theory evaluation  of the BPS correlator. 
Our result for the semiclassical trajectory 
 will agree with the  3-geodesic intersection in \ci{km} in its $AdS_5$ 
 part 
 (but its  $S^5$ part will be different from that suggested in \ci{km}). 
Another  important ingredient of our construction will
 be an analog of the Schwarz-Christoffel map 
used in the light-cone  interacting string  diagrams in flat space \ci{man}
with $\D_i$ (or $J_i$ in the BPS case) playing the role of
 light-cone momenta $p^+_i$ 
determining string lengths.
It    will be used to construct the semiclassical 
solution by first mapping the complex plane with 3 marked points   corresponding to the  three 
 vertex operator insertions 
 to a domain
in $\tau + i \sigma $ 
plane  (which generalizes the  usual   cylinder in  the 2-point function  case)
and then choosing the simplest ``point-like'' solution  which is 
 linear in $\tau$.  
An   important   difference compared   to the  flat space case (where $p^+$ is conserved) 
will be ``non-conservation'' of $\D_i$ for  non-extremal 
correlators.\foot{In the extremal  correlator case the 
map will be same as in the flat space --
 describing one cylinder becoming two joined cylinders 
  with the sum of the two lengths matching the length of the original cylinder.}
To construct the corresponding 
  map for arbitrary values of  $\D_i$ we will  start with  a
generic expression for the $AdS_5$ string stress tensor 
 having  prescribed singularities  at  
the punctures; its form is uniquely determined by $\D_i$ as was pointed out 
 in \ci{ja}.

For non-BPS  operators   with  large quantum numbers in $S^5$ only, i.e. 
describing semiclassical strings  ``extended'' only in $S^5$,   one may expect 
that the $AdS_5$ part  of the semiclassical trajectory should be  the same 
as in the  case of the BPS correlator with  generic non-extremal
  choice of the dimensions  $\D_i$. 
Given that in the conformal gauge  the $AdS_5$ and $S^5$ parts of the string equations for 
the  semiclassical trajectory decouple, 
  looking only at  the $AdS_5$ part of the semiclassical trajectory
   one should  not  be able to see  the difference 
between  the cases of a    non-BPS  correlator 
(with non-trivial parts of vertex operators depending only on  $S^5$ coordinates) 
 and a BPS one with the same dimensions $\D_i$. 
 
 This is  the point of view we shall 
 try to justify in this paper. 
 At the same time,  the picture  suggested recently in  \ci{ja}
  is different: it was argued there that 
 for generic $\D_i$ 
 the   semiclassical solution  should be extended in $AdS_2$, 
 becoming  approximately point-like as  in  \ci{km} only for sufficiently
  small $d_i= {\D_i \ov \sql}$. 
 This proposal, however, raises few questions.
 The BPS  case should be a special limit
  of a non-BPS case  but as the $AdS_5$ part of the solution depends 
  only on $\D_i$ there is no way 
  to  tell the difference between the two.
   Also, there is no  natural ``scale''
  to compare  $d_i $ to,   so the  notion  of correspondence  with the BPS case  only 
  ``for sufficiently small'' $d_i$    seems  artificial, given  that the 
  BPS states can  carry  arbitrarily  large 
  charges/dimensions.\foot{A technical reason  for this ``smallness'' condition in \ci{ja}
  appears to be as follows.  The string solution in \ci{ja} is constructed 
  from a  non-linear generalized sinh-Gordon equation 
  equation $\del \bar \del \td \g = \sqrt{ T \bar T } \sinh \td \g $ where 
  $T$ is an effective  2d stress tensor that  scales as $d^2_i= { \D^2_i \ov (\sql)^2}$.
   Thus for small 
  $d_i$ the solution 
   approximates to  $\td \g=0$ solution    which indeed corresponds to a point-like string. 
   At the same  time, $\td \g=0$ solution   exists  even for an  arbitrarily large $d_i$.
   This is, in fact, the  choice  that we will advocate here. 
   More general solutions that appear to  
   represent
   surfaces extended in $AdS_5$ appear to represent 
   states that carry extra hidden $AdS_5$ charges and 
   thus are more general
  than the  states with only $S^5$  charges.
  }

 We shall start in  section 2   with a discussion of the supergravity 
  representation for  the  protected  3-point function of BPS states 
  given by  an integral of a product 
  of the three  bulk-to-boundary propagators  and three ``spherical harmonic'' 
factors  over a point 
  of $AdS_5\times  S^5$.  We will show 
  that in  the limit when the 
  dimensions $\D_i$ are large   this integral   is saturated by a stationary point. 
  In the  $AdS_5$ part 
  this point is the same as found in \ci{km}. 
We   will  show that for 
 a non-extremal
  correlator  the stationary point for the  $S^5$ 
  part of the integral can be found in a similar way by using  an analytic continuation
 trick   relating the $S^5$ problem to an effective  $AdS_5$ one. 
  We will also prove    that the  resulting expression for the 
  large charge limit of the correlator agrees, as expected, 
   with the one found on  the free gauge theory side.

In section 3   we shall consider the \adss  string-theory representation for the 2-point 
and 3-point
functions  in terms of correlators of the  corresponding  marginal 
vertex operators following \ci{bt1,bt2}.
 In section 3.1  we shall clarify the issue
of cancellation of volumes of residual  world-sheet and $AdS$ target space 
conformal transformations  leading to finite expressions for the 2-point and 3-point functions. 
In section 3.2 we shall  review the semiclassical  approximation 
 for the 2-point function of  BPS operators with large charges.

In section 4 we shall study the  semiclassical approximation  for the  string theory 
representation of
 extremal ($\D_1=\D_2+\D_3$) correlator of the  3 BPS operators with large charges
 and show that the corresponding semiclassical trajectory can  be interpreted as 
 an intersection of 3 euclidean $AdS_5$ geodesics  as suggested in \ci{km}.
 We will demonstrate 
  that this   interpretation applies  provided one  first  maps the complex 
  plane with 3  punctures into a
 cylindrical domain by the same  Schwarz-Christoffel map as in the flat-space 
 light-cone interacting string
 picture. This map encodes the positions of insertions of the vertex operators
  on the complex plane.

In section 5 we will generalize   to non-extremal 3-point correlator   case. 
Our discussion of  the $AdS$ part of the solution in section 5.1 
will be  completely general, i.e.  applicable to  all (BPS or non-BPS) 
 states with large  quantum numbers only  in $S^5$. 
 We  will show that 
the  $AdS$  solution is still 
given by the  3 appropriately glued geodesics but
 the transformation to the complex plane is now 
given by a more general 
Schwarz--Christoffel map  (which corresponds to the case
 of ``non-conservation'' of  string lengths or $p^+$ in the 
corresponding flat space case). 
The precise form of the Schwarz--Christoffel map is dictated by the $AdS$ 
stress  tensor. In section 5.2 we will specify to the case of 
 non-extremal BPS correlator.
 Guided  by the supergravity  discussion in section 2 we will 
find the corresponding semiclassical trajectory in $S^5$  using 
an analytic continuation  to $AdS_5$   and finally show that we get the 
same expected semiclassical expression for the correlator  as in the 
supergravity  approximation.

In section 6 we  will consider a
particular example  of  a 3-point correlation  function of non-BPS operators
representing  ``small'' 
circular strings with two equal spins  in $S^3\subset S^5$. 
 In the case when all the three operators
 represent states in the same $S^3$ of $S^5$ we will find a contradiction between 
 the angular momentum conservation condition and 
 the non-linear  on-shell (i.e. marginality) condition $\D^2_i= 4 \sql {J_i}$
 suggesting that this correlator should vanish as in  the corresponding flat space case.  

We will conclude in section 7  with some 
comments  on a comparison of  our approach (of section 5.1) 
to the construction of generic $AdS$ solution 
with  that of   ref. \cite{ja}.
The  solution   constructed in~\cite{ja} is  more general 
than ours, but it is not clear if it is actually  necessary
to describe the  correlators   of states    with non-trivial charges  in $S^5$ only.
As we will argue, 
the relevant $AdS_5$  solution should  be  the ``point-like'' one of section 5.1
that should  universally apply to both BPS and  non-BPS cases. 

In Appendix A we will  elaborate  on the issue of cancellation of
the  Mobius group volume factor and 
 the volume of the 
 residual $AdS_5$  symmetry group   transformations in the 2-point and 
 3-point correlators of string vertex operators. 
 In Appendix B we will  give details of the Schwarz-Christoffel map 
 constructed  in section 5.


\section{Semiclassical three-point functions in supergravity}


In this paper we will study 3-point functions of ``heavy'' scalar 
operators whose dimensions 
$\D_i$, $i=1, 2, 3$ scale as $\D_i \sim \sqrt{\lambda}$ for large 't Hooft coupling 
$\lambda$. We will be interested only in the leading semiclassical contribution
of order $e^{a\sqrt{\lambda}}$, i.e. will be  ignoring  subleading corrections. 
For this reason it will be possible to ignore  detailed structure of the corresponding 
vertex operators or  wave functions. 

In this section we will 
consider the calculation of such  3-point function in supergravity. 
By  semiclassical approximation here we shall 
assume the  limit of large dimensions or charges in which the 
\adss integral will be saturated by a stationary  point approximation. 
While the full  calculation will  be  valid 
for   BPS states only, 
 the  $AdS_5$  part of 
 it will 
formally apply also  to the case of operators representing 
semiclassical string states  that 
do not carry other $AdS_5$  quantum numbers  except the energy: 
  they will be described by an effective $AdS_5$ action with a local cubic interaction.

 In supergravity  description the 3-point function 
 is given by  a simple Witten 
diagram consisting of  three bulk-to-boundary propagators as in 
Figure 1~\cite{Witten98,Seiberg}.
\begin{figure}[ht]
\centering
\includegraphics[width=40mm]{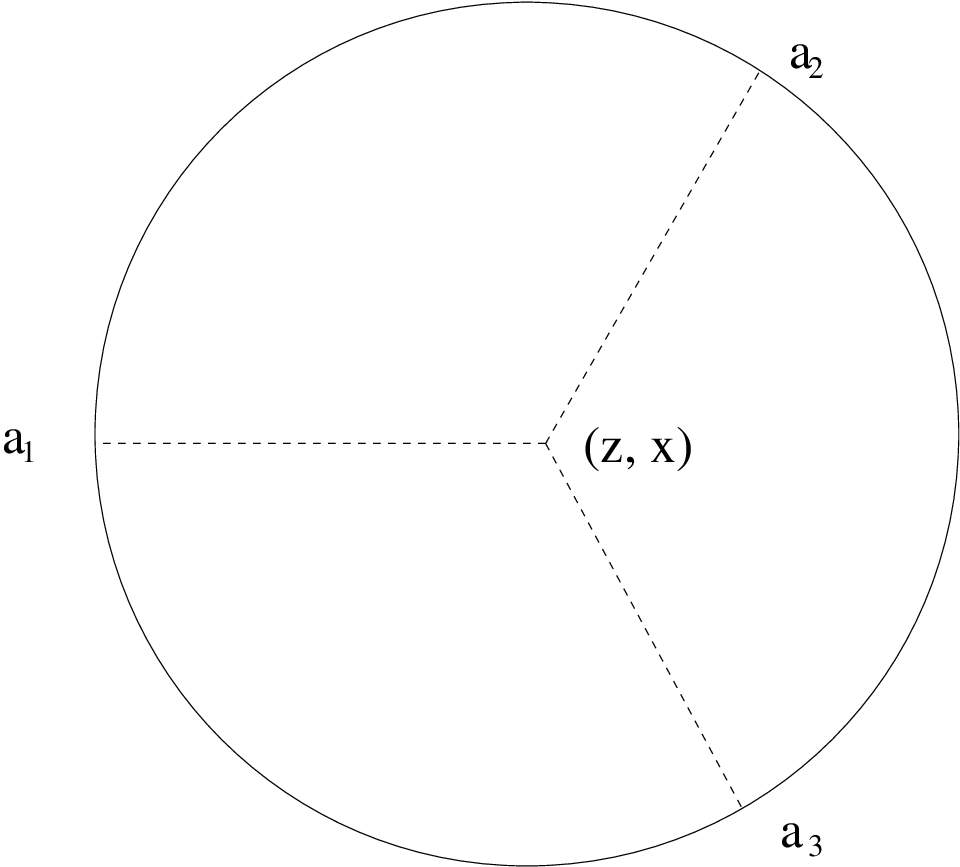}
\caption{\small Witten diagram for  three-point function in supergravity.
}
\label{fig1}
\end{figure}
The contribution of this diagram splits into the product of the $AdS_5$-factor
and the $S^5$-factor:\foot{As already mentioned,  as 
 we are interesting in the leading semiclassical (large dimension/charge)
limit of the correlator it is sufficient to ignore  details  of factors 
in the integrands  that do not scale as powers of $\D_i$ or $J_i$.}
\be
G = G_{AdS}(\vec{a}_1, \vec{a}_2, \vec{a}_3)\ 
 G_{S^5}(n_1, n_2, n_3) \,, 
\label{1.1}
\ee
where\foot{ 
 In global coordinate parametrization (see section 3.3)
 the $S^5$ integral reduces to a gaussian integral and 
 Gamma-function  factor resulting from the integral 
 over the Lagrange  multiplier. The $AdS_5$ integral can be 
 defined by an analytic
 continuation. 
 Expressions for 3-point integrals like
 this one or its analog in $S^5$ 
(cf. section 2.4) were computed  also  
 in \ci{br}.}
\be
G_{AdS}(\vec{a}_1, \vec{a}_2, \vec{a}_3) \sim 
\int\frac{d^4x dz}{z^5}\ [K(\vec{a}_1)]^{\D_1}
[K(\vec{a}_2)]^{\D_2}[K(\vec{a}_3)]^{\D_3} \ , 
\label{1.2}
\ee
and
\be
G_{S^5}(n_1, n_2, n_3) \sim \int d \Omega {\ } U_1^{J_1} U_2^{J_2} U_3^{J_3}\,.
\label{1.3} 
\ee
%
Here we consider euclidean  $AdS_5$ in the Poincare coordinates with the metric
%
\be
d s^2 =\frac{1}{z^2}(d z^2 + d \vec{x}^2)\,, \qquad \vec{x}=x^m=(x_0, x_1, x_2, x_3)\,,
\label{1.5}
\ee
and 
\be
K(\vec{a}_i) =\frac{z}{z^2+ (\vec{x} -\vec{a}_i)^2}\,.
\label{1.4}
\ee
In the integral over $S^5$ in~\eqref{1.3} the functions 
$U_i$  \ ($i=1,2,3$) specify the 
  three 
states   under consideration. In general, they can be written as
\be
U_i =n_i \cdot  X =   \sum_{p=1}^6 n_{i p} X_p\,, \quad \quad \quad \sum_{p=1}^6 X_p^2 =1\,,
\label{1.5.5}
\ee
where the  complex 6-vectors $n_i$  are constrained to satisfy 
\be 
n_i \cdot n_i =0\,, \quad  \quad  \quad n_i \cdot n_i^* =2\,.
\label{1.6}
\ee
%
 Note that for the BPS states
we must have $\Delta_i =J_i$ for large charges $J_i$. 

On general grounds of $SO(2,4) \times SO(6)$   invariance we should expect that $G$ in 
\rf{1.1}  should have the following structure ($\a_1= \Delta_2+\Delta_3-\Delta_1$, etc.)
\be
G 
=\frac{C}{|\vec{a}_1- \vec{a}_2|^{\a_3}|\vec{a}_1- \vec{a}_3|^{\a_2}|\vec{a}_2- \vec{a}_3|^{\a_1}}\,,
\label{1.21.00}
\ee
where the coefficient $C$ should  be  a function of  the scalar products $ n_i \cdot n_j$  (i.e.
$C=C(n_1 \cdot n_2, n_2 \cdot n_3, n_3 \cdot n_1)$) and also of the quantum numbers 
$\D_i=J_i$.



\subsection{Two-point function}

It is useful  to  review  first   the case of the 
 2-point function (see Appendix B in ~\cite{bt2}). 
It is given by the same expressions as in ~\eqref{1.1},\eqref{1.2},\eqref{1.3}
with $\D_3=0, \ 
\D_1=\D_2=\D$, $J_1=J_2=J$ and $U_2=U_1^*$ (i.e. $n_2=n_1^*$).  
The $AdS_5$ contribution is then
\be
G_{AdS}(\vec{a}_1, \vec{a}_2) \sim \int \frac{d^4 x d z}{z^5} \ 
[K(\vec{a}_1)]^{\D} [K(\vec{a}_2)]^{\D}\,.
\label{1.7}
\ee
In the limit of large $\Delta$ this  integral is saturated by the stationary 
point of the effective 
 ``action''
\be
-A_{AdS}=  \D \ln \frac{z}{z^2+ (\vec{x} -\vec{a}_1)^2} +
\D \ln \frac{z}{z^2+ (\vec{x} -\vec{a}_2)^2}\,.
\label{1.8}
\ee
Without loss of generality we can choose $\vec{a}_1$, $\vec{a}_2$ to lie
along the $x_0$ (euclidean time) axis, $\vec{a}_i =(a_i, 0, 0, 0)$ and set $a_1=0$, 
$a_2\equiv a >0$. 
Then  we can  choose solution with 
 $x_1=x_2=x_3=0$, i.e. $ \vec{x}=(x_0,0,0,0)$, 
 so that  the  only  non-trivial 
equations  are obtained by varying~\eqref{1.8} with respect to $x_0$ and $z$
(for notational simplicity we will ignore the subscript ``0'', i.e. set 
 $x_0\equiv x$)
\bea
\frac{x}{z^2+ x^2} + \frac{x-a}{z^2+ (x-a)^2}=0\,, \ \ \ \ \ \ \ \ \ \ \ 
\frac{z^2- x^2}{z^2+ x^2} + \frac{z^2 -(x-a)^2}{z^2+ (x-a)^2}=0\,.
\label{1.9}
\eea
The solution of these equations  found in~\cite{yo, bt2}  is given by a half-circle 
in $(x,z)$ half-plane:
\be
z^2= x (a-x)
\label{1.10}
\ee
or, equivalently, 
\be 
z=\frac{a}{2 \cosh \tau}\,, \quad  \quad  \quad 
x=\frac{a}{2} \tanh \tau + \frac{a}{2}\,.
\label{1.11}
\ee
This line is a geodesic is $AdS_2 \subset AdS_5$ connecting the boundary points 
$x=0$ and $x=a$. Evaluating~\eqref{1.7} on this solution gives
\be
G_{AdS} \sim \frac{1}{a^{2 \D}} \int_{-\infty}^{\infty} d \tau\  Q^{-1/2}(\tau)\,, 
\label{1.12}
\ee
where $Q(\tau)$  is the 
 ``one-loop'' determinant of small  fluctuation operator 
around the solution~\eqref{1.11}. This integral over $\tau$ gives an 
order 1 correction that we ignore here. 

The integral over $S^5$ is 
\be
G_{S^5} \sim \int d \Omega \  (n_{p} X_p)^{J_1} (n_{p}^* X_p)^{J_2}\,.
\label{1.13}
\ee
We can always choose the coordinates on $S^5$ so that 
\be
n_p X_p = \cos \psi\ e^{i \varphi}\,, \quad \quad \quad 
n_p^* X_p = \cos \psi\ e^{-i \varphi}\,.
\label{1.14}
\ee
Then 
the   integral over $\varphi$ implies  charge conservation $J_1=J_2$
and for large $J_i$ the integral over $\psi$ is saturated by 
$\psi=0$. Then 
$G_{S^5} \sim 1$. 
Combining the $AdS_5$ \rf{1.12} and $S^5$ \rf{1.13} parts together gives 
\be 
G(a_1=0, a_2=a) \sim \frac{1}{a^{2 \D}}
\label{1.15}
\ee
up to terms of order unity. We have thus 
  obtained the 2-point 
function  which is canonically normalized up to terms that are subleading for  $\D \gg 1$.


\subsection{$AdS_5$-contribution to 3-point function}


In a similar way, in  the limit of large $\Delta_i$'s the
 integral~\eqref{1.2}  can be evaluated 
by extremizing the ``action''
\be
-A_{AdS}= \D_1 \ln \frac{z}{z^2+ (\vec{x} -\vec{a}_1)^2} +
\D_2 \ln \frac{z}{z^2+ (\vec{x} -\vec{a}_2)^2}
+\D_3 \ln \frac{z}{z^2+ (\vec{x} -\vec{a}_3)^2}\,.
\label{1.16}
\ee
Again, without loss of generality we can choose the 3 points to lie along the $x_0 $-axis, 
i.e. $\vec{a}_i= (a_i, 0, 0, 0)$,  and set $a_1=0$,\  $0 < a_2 <a_3$. 
Then it follows from~\eqref{1.16} that the equations for $x_1,x_2,x_3$ are 
satisfied by $x_1=x_2=x_3=0$ and the 
remaining  non-trivial equations for $x\equiv x_0$ and $z$ take the form 
\bea
&&
\D_1\frac{ x}{z^2+ x^2} + \D_2 \frac{x-a_2}{z^2+ (x-a_2)^2}
+  \D_3 \frac{x-a_3}{z^2+ (x-a_3)^2} =0\,, \nonumber \\
&&
\D_1 \frac{z^2- x^2}{z^2+ x^2} + \D_2 \frac{z^2 -(x-a_2)^2}{z^2+ (x-a_2)^2}
+ \D_3 \frac{z^2 -(x-a_3)^2}{z^2+ (x-a_3)^2}
=0
\label{1.17}
\eea
The solution to eqs.~\eqref{1.17} was found in~\cite{km}  and 
 is given by an isolated ``interaction'' point 
\bea
&&
x_{int} = \frac{\a_1 a_2  a_3 (\a_2 a_2 +  \a_3  a_3) }{\a_1 \a_2 a_2^2 +\a_1 \a_3 a_3^2
+\a_2 \a_3 (a_3-a_2)^2} \,, \nonumber \\
&&
z_{int} =\frac{\sqrt{ \a_1 \a_2 \a_3 (\a_1 +\a_2 +\a_3)} (a_3-a_2) a_2 a_3}{\a_1 \a_2 a_2^2 +\a_1 \a_3 a_3^2
+\a_2 \a_3 (a_3-a_2)^2} \,, 
\label{1.18}
\eea
where 
\bea
\a_1= \D_2+\D_3-\D_1\,, 
\ \ \ \ \ \ \ \ 
\a_2= \D_1+\D_3-\D_2\,, 
\ \ \ \ \ \ \ \ 
\a_3= \D_1+\D_2-\D_3\,. 
\label{1.19}
\eea
Note that if the correlator is extremal,
i.e.  $\D_1=\D_2+\D_3$, then $\a_1=0$ 
and the extremum \rf{1.18} lies on the boundary ($z=0$).
 This leads to a  divergence of the 
``action''~\eqref{1.16}. In this case  
 the  correlator should be defined as a limit of non-extremal one, i.e. 
by starting with   $\D_1=\D_2+\D_3 + \eps $ and taking $\a_1 = \eps \to 0 $
 at the very end. 
 

Evaluating~\eqref{1.16} on the solution \rf{1.18} leads to the following
semiclassical approximation to the $AdS$ part of the 3-point correlator \rf{1.2}
\cite{km}
\bea
&&G_{AdS} (a_1=0, a_2, a_3) \sim \frac{C_{AdS}}{a_2^{\a_3} a_3^{\a_2} (a_2-a_3)^{\a_1}}\,, 
\label{1.20} \\
%
%
&&C_{AdS} = \Big[ \frac{\a_1^{\a_1} \a_2^{\a_2}  \a_3^{\a_3} 
(\a_1+\a_2+\a_3)^{\a_1+\a_2+\a_3}}{(\a_1+\a_2)^{\a_1+\a_2} (\a_1+\a_3)^{\a_1+\a_3}
 (\a_2+\a_3)^{\a_2+\a_3}} \Big]^{1/2}\,.
\label{1.21}
\eea
The resulting dependence on space-time points $\vec{a}_i$  is consistent with 
 conformal  invariance, implying that in general 
\be
G_{AdS} (\vec{a}_1, \vec{a}_2, \vec{a}_3)
=\frac{C_{AdS}}{|\vec{a}_1- \vec{a}_2|^{\a_3}|\vec{a}_1- \vec{a}_3|^{\a_2}|\vec{a}_2- \vec{a}_3|^{\a_1}}\,.
\label{1.21.0}
\ee
%
Note also that in the extremal limit $\a_1 \to 0$ the expression in eq.~\eqref{1.21} 
is finite and gives $C_{AdS}=1$.


\subsection{$S^5$-contribution to 3-point function}


In the extremal case    $\D_1=\D_2+\D_3$       we  may  choose 
\be 
n_2=n_3=n_1^*\ ,
\label{1.21.1}  
\ee
and then the semiclassical evaluation  of the integral over $S^5$ in ~\eqref{1.3}
is  similar to that for the 2-point function 
and produces the contribution  $C_{S^5}\sim 1$. 

In  the non-extremal case  it is useful first to consider  a particular example   
and then present a generalization to the case of arbitrary 
 complex 6-vectors $n_i$ subject to \rf{1.6} in the  next subsection. 
Namely, let us choose $n_i$ as 
\be
n_1=(1, i, 0, 0, 0, 0)\,, \qquad 
n_2=(1, -i, 0, 0, 0, 0)\,, \qquad 
n_3=(1, 0, i, 0, 0, 0)\ , 
\label{1.22}
\ee
 corresponding  to
\be
U_1= X_1+ i X_2 \,, \qquad U_2=X_1-i X_2\,, \qquad 
U_3=X_1+i  X_3\,.
\label{1.23}
\ee
Let us  parametrize  the metric on $S^5$ as 
\be
d s^2 = d \theta^2 + \sin^2 \theta d \varphi_3^2 +\cos^2 \theta
(\cos^2 \psi d\varphi_1^2 + \sin^2 \psi d\varphi_2^2 )\,.
\label{1.23.1}
\ee
The choice of $U_i$'s in \rf{1.23} effectively allows to reduce the problem to $S^2$
i.e. (below $\varphi_1\equiv \varphi$)
\bea
&& \theta =0\,, \qquad \varphi_2=0\,, \qquad \varphi_3=0\,, \nonumber \\
&& X_1+i X_2= \cos \psi\ e^{i \varphi}\,, \qquad
 X_1-i X_2= \cos \psi\ e^{-i \varphi}\,, \qquad X_3=\sin \psi\,.
\label{1.24}
\eea
It is useful to  consider the following analytic continuation
%
\be
i X_2 \to \tX_2\,, \qquad i X_3 \to \tX_3\,, \qquad i \varphi \to \tvarphi\,, \qquad i \psi
\to \tpsi\,.
\label{1.25}
\ee
We will see that the extremum is real in these  ``rotated'' coordinates. 
The  effective  action for $S^5$ integral \rf{1.3} then  becomes
\be 
-A_{S^5}=(J_1+ J_2) \ln (\cosh \tpsi) + (J_1-J_2) \tvarphi 
+ J_3 \ln (\cosh \tpsi \cosh \tvarphi +\sinh \tpsi)\,.
\label{1.26}
\ee
Varying it with respect to $\psi$ and $\varphi$ gives
\bea
&& (J_1+J_2) \tanh \tpsi +\frac{J_3 (\sinh \tpsi \cosh \tvarphi+
 \cosh \tpsi)}{\cosh \tpsi \cosh \tvarphi+ \sinh \tpsi}=0\,, 
\no \\ && \ \ \ 
J_1-J_2 + \frac{J_3 \cosh \tpsi \sinh \tvarphi}{\cosh \tpsi \cosh \tvarphi+ \sinh \tpsi}=0 \ . 
\label{1.27}
\eea
One can write  
the solution to these two equations in the following form
\bea
\tanh \tvarphi_{int} =\frac{(J_1-J_2) (J_1+J_2-J_3)}{(J_1-J_2)^2 - (J_1+J_2)J_3}\,, 
\ \ \ \ \ \ \ \ 
\tanh \tpsi_{int} =\frac{1}{2} \sqrt{\frac{J_3^2 - (J_1-J_2)^2}{J_1 J_2}}\,.
\label{1.28}
\eea
Evaluating $U_1$,$U_2$,$U_3$ on this solution gives
\bea
&&
U_1= X_1+\tX_2 =\frac{(\beta_2+\b_3)\sqrt{\b_1}}{\sqrt{\b_2 \b_3 (\b_1+\b_2+\b_3)}}\,, 
\ \ \ \ \ \ \ \ 
U_2 =X_1-\tX_2 =\frac{(\beta_1+\b_3)\sqrt{\b_2}}{\sqrt{\b_1 \b_3 (\b_1+\b_2+\b_3)}}\,, 
\nonumber \\
&&\ \ \ \ \ \ \ \ 
U_3 = X_1+\tX_3 = \frac{1}{2}
\frac{(\beta_1+\b_2)\sqrt{\b_3}}{\sqrt{\b_1 \b_2 (\b_1+\b_2+\b_3)}}\,, 
\label{1.29}
\eea
where $\beta_i$ are given  by  similar expressions as $\alpha_i$ 
in \rf{1.19} with $\D_i \to J_i$ 
\be
\b_1= J_2+J_3-J_1\,, \qquad 
\b_2= J_1+J_3-J_2\,, \qquad 
\b_3= J_1+J_2-J_3\,.
\label{1.30}
\ee
Then the stationary-point value of the $S^5$ 
integral~\eqref{1.3}  is  found to be 
\be
G_{S^5} \equiv C_{S^5} \sim \frac{1}{2^{J_3}}\ 
\Big[ \frac{ (\b_1+\b_2)^{\b_1+\b_2} (\b_1+\b_3)^{\b_1+\b_3} (\b_2+\b_3)^{\b_2+\b_3}}
{\b_1^{\b_1}\b_2^{\b_2} \b_2^{\b_2} (\b_1+\b_2+\b_3)^{\b_1+\b_2+\b_3}}\Big]^{1/2}\,.
\label{1.31}
\ee
Combining it with the $AdS_5$ part  eq.~\eqref{1.21} and using that 
$\D_i= J_i$ (i.e.  $\a_i=\b_i$) we 
find  that the $S^5$-contribution almost 
completely cancels the contribution from $AdS_5$:  
 up to subleading terms   we find for 
 the 3-point  coefficient $C$ in \rf{1.21.00}\foot{The asymmetry of this expression in
 $J_i$ has, of course, to do with our particular choice of $U_i$ in \rf{1.23}.}
\be
C= C_{AdS}\ C_{S^5} =\frac{1}{2^{J_3}}\,.
\label{1.32}
\ee
It is  useful  to rederive this result 
in a different way   that explains 
 why this near-cancellation between the 
$AdS_5$ and $S^5$ parts  happens. 
When we perform the analytic continuation~\eqref{1.25} we effectively 
turn $S^2 \subset S^5$ into the euclidean $AdS_2$ 
defined  by $X_1^2 -\tX_2^2-\tX_3^2 =1$. 
In this $AdS_2$ space we  may  introduce the Poincare coordinates $(r, y)$  as 
\be
X_1= \frac{1}{2r} (1+r^2 +y^2)\,, \qquad 
\tX_2 =\frac{y}{r}\,, \qquad \tX_3=\frac{1}{2r}(-1+r^2+y^2)\,.
\label{1.32.1}
\ee
Then the $U_i$ in~\eqref{1.23},\rf{1.25}  become
\bea
U_1=\frac{1}{2}\Big(\frac{r}{r^2+ (y+1)^2}\Big)^{-1}\,, 
\  \ \ \ \ \ \ \ 
U_2=\frac{1}{2}\Big(\frac{r}{r^2+ (y-1)^2}\Big)^{-1}\,, 
\  \ \ \ \ \ \ \ 
U_3=\Big(\frac{r}{r^2+ y^2}\Big)^{-1}\,. 
\label{1.32.2}
\eea
These expressions look   the same -- up to $1\ov 2$ factors and inverse powers -- 
as  the 
 bulk-to-boundary propagators~\eqref{1.4} in $AdS_2$ 
where the boundary points are  chosen as -1, 1, 0. 
This implies that in evaluating 
integral over the sphere~\eqref{1.3} in the stationary-point approximation 
we can immediately borrow the $AdS_5$ result~ \rf{1.20}--\eqref{1.21.0} 
in which we should  substitute\footnote{The semiclassical 
solution on $S^5$~\eqref{1.28} is exactly the same as 
its $AdS$ counterpart~\eqref{1.18} 
with the following replacements:
 $a_2 \to 1,\ a_3 \to -1,\ \a_1 \to -\b_2,\ \a_2 \to
 -\b_1,\ \a_3 \to -\b_3$.
This  can be easily 
verified using eqs.~\eqref{1.32.1} and~\eqref{1.29}.}
\be
a_1 \to -1\,, \qquad a_2 \to 1\,, \qquad a_3 \to 0\,, \qquad 
\a_i \to -\b_i\,.
\label{1.32.3}
\ee
Here $\a_i \to -\b_i$ is due to the negative powers   in~\eqref{1.32.2}.
This is the very reason why 
the  above cancellation between the $AdS_5$ and $S^5$ contributions 
takes place. 
Taking into account  the factors $\frac{1}{2}$ in $U_1$ and $U_2$ in \rf{1.32.2}
we get  $2^{-J_1-J_2}$; the factor 
${|\vec{a}_1- \vec{a}_2|^{\b_3}|\vec{a}_1- \vec{a}_3|^{\b_2}|\vec{a}_2- \vec{a}_3|^{\b_1}}\,$
gives  $2^{\b_3}$.
Combining these  together we end up again  with ~\eqref{1.31},\rf{1.32}.

Let us note that in the special case 
when one of the dimensions vanish,  $\D_3=J_3=0$, 
the 3-point function \rf{1.21.00},\rf{1.32} reduces to the 2-point one \rf{1.15}
if $\D_1=\D_2$. 

\subsection{Generic  non-extremal 3-point  function }

Let us now generalize the analytic continuation 
 trick used  at the end of  the previous section 
to compute the semiclassical expression for \rf{1.1}  for a more general 
 choice of 6-vectors in \rf{1.5.5},\rf{1.6} which allows 
 to restore  the full expression for the coefficient $C$. 
 
Let  us  start with the euclidean $AdS_5$ space  and  introduce 6 embedding coordinates
$Y_r$ \ ($r=(-1, m,4); \ m=0,1,2,3$) related to the Poincare coordinates in \rf{1.5}  as 
\bea
&&Y_{-1} = \frac{1}{2z} (1+z^2 +\vec x^2)\,, \qquad 
Y_m   =\frac{x_m }{z}\,, \qquad   Y_4=\frac{1}{2z}(-1+z^2+\vec x ^2)\,, \la{01}\\
&&   (Y, Y)  \equiv   Y_{-1}^2 -   Y_m Y_m - Y_4 Y_4  =1  \ .
\eea
Then it is easy to check that  the inverse of the bulk-to-boundary 
propagator 
$K(\vec b)$\footnote{Since the $AdS_5$ space discussed in this 
 subsection  will play an auxiliary role, 
 we will denote the boundary points by   ${\vec b}_i$ rather than by ${\vec a}_i$.
} 
in \rf{1.4}  can be written as a linear combination of  $Y_r$ 
\bea 
&&  [K(\vec b)]^{-1}= (\nN,Y)  \ , \ \ \ \ \  \ \ \ \ \ \ 
 \nN= ( 1+ \vec b^2, - 2 \vec b ,  1- \vec b^2) \equiv  (1 +  \vec b^2)\ \hat n \ , 
 \la{02} \\
&&   (\hat  n ,\hat n ) =0 \ , \ \ \ \ \ \ \ \ \  \ \  \hat n  \cdot \hat n  \equiv \sum_p 
\hat n_r \hat n_r =  2 \ . \la{03}
\eea
This shows that we can equivalently parametrize the bulk-to-boundary 
propagator $K(\vec b)$ in terms of the vector $\hat{n}$.
In particular, for $\vec b=0$, 
we  have $\nN= ( 1, 0,0,0, 0 ,  1) , \  \ [K(0)]^\D = (Y_{-1} + Y_4)^{-\D}$.
Analytically continuing  to $S^5$
\be 
Y_{-1}= X_5 \ , \ \ \ \ \   Y_m = i X_m \ , \ \ \ \ \  Y_4 = i X_4 \ , \ \ \ \ \ \ \ \    
 X_p X_p =1  \ ,  \la{04} \ee
  and introducing a complex 6-vector
 $n=(\hat n_1, i \hat n_m, i \hat n_4)$ satisfying \rf{1.6} as 
\be
 n=  \big( 1, - i {2 \vec  b\ov 1 + \vec b^2} , 
  i {1- \vec b^2\ov 1+ \vec b^2}  \big) \ , \ \ \ \ \ \ \ \ 
n \cdot n =  (\hat  n ,\hat n ) =0 \ , \ \ \ \ \ \  n \cdot n^* =
  \hat n  \cdot \hat n =2\ , 
 \la{05}
\ee
we get  
\be     [K(\vec b)]^{-1}= (\nN,Y) =   (1 +  \vec b^2)\  n \cdot  X \ , \la{06} \ee
and thus  find a map between the semiclassical $S^5$ problem  in \rf{1.1}, \rf{1.3}
and an equivalent $AdS_5$ problem. 

Now let us consider three generic states in $S^5$ of the form~\eqref{1.5.5}, \eqref{1.6}.
Each vector $n_i$, $i=1,2,3$  contains 
 $2 \times 6-3=9$ independent real parameters\footnote{The moduli 
space of a single geodesic can be viewed as  8-dimensional Grassmanian  
$SO(6)/[ SO(4) \times SO(2)]$ (see, e.g., \ci{mi}): 
in addition to the real 
$SO(6)$ invariance of the two constraints \rf{1.6}, they are also invariant under
a multiplication of $n$ by a phase.}
so that overall the three states
are characterized by $3 \times 9=27$ real parameters. In addition, we are allowed to act on $n_i$ with
 $SO(6)$
transformations preserving~\eqref{1.6}. We can use this $SO(6)$ freedom 
to restrict the number of independent real parameters to 
27-dim $SO(6)$= 12=4$\cdot$3. Hence, we can always choose
each of the three vectors $n_i$ in the form~\eqref{05}, i.e.
  parametrized by a real 4-vector $\vec b_i$.   
Then 
\be 
 G_{S^5} ( n_1,n_2,n_3)= \bra \prod_{i=1}^3 (n_i \cdot  X)^{J_i}  \ket_{_{S^5}  }
= \prod_{k=1}^3  (1 +  \vec b^2_k)^{-J_k} \ \  \bra  \prod_{i=1}^3    [K(\vec b_i)]^{-J_i} \ket_{_{AdS} } \ . \la{07}
\ee
Since here $-J_i$ appear in place of $\D_i$  in \rf{1.2} 
that   means that we  get the same semiclassical trajectory as 
in  the original $AdS_5$ case 
with $\a_i \to \b_i$ in 
\rf{1.30} but  the  total contribution should 
 appear in the opposite power.
We then find 
\bea 
 G_{S^5} ( n_1,n_2,n_3)&=&\prod_{k=1}^3  (1 + \vec b^2_k)^{-J_k}\ \
 G_{AdS} (\vec{b}_1, \vec{b}_2, \vec{b}_3)_{_{ \D_i \to - J_i }} \no\\
&=&
\prod_{k=1}^3  (1 + \vec b^2_k)^{-J_k}
\ {C^{-1}_{AdS}(\b_i) }\ {|\vec{b}_1- 
\vec{b}_2|^{\b_3}\ |\vec{b}_1- \vec{b}_3|^{\b_2}\ |\vec{b}_2- \vec{b}_3 |^{\b_1}}\,. \label{08}
\eea
Observing that  for $n_i$ defined as in  \rf{05} 
we have 
\be 
n_1 \cdot n_2 = {   2 (\vec b_1 - \vec b_2)^2 \ov  (1 +  \vec b_1^2) 
 (1 + \vec b_2^2) } \ , 
\la{09}
\ee
we can then rewrite \rf{08} in terms of $n_i$   as
\be 
 G_{S^5} ( n_1,n_2,n_3) = {C^{-1}_{AdS}(\b_i) }\  \big({n_1 \cdot n_2\ov 2}\big)^{\b_3 /2} 
\big({n_2 \cdot n_3\ov 2}\big)^{\b_1 /2}   \big({n_1 \cdot n_3\ov 2}\big)^{\b_2 /2}   \ . \la{010}
\ee 
Finally, using  \rf{1.1}, we get  the following expression   for the  coefficient $C$ in the 
 full semiclassical  \adss  correlator \rf{1.21.00}
\bea 
&&C= (\tilde{n}_1 \cdot \tilde{n}_2)^{\b_3/2}\ (\tilde{n}_1 \cdot \tilde{n}_3)^{\b_2/2}\  (\tilde{n}_2 \cdot \tilde{n}_3)^{\b_1/2}\,,
\label{011}\\
&& \tilde{n}_i \equiv  {\textstyle{1 \ov \sqrt 2}} n_i \ , \ \ \ \ \ \ \
\tilde{n} \cdot \tilde{n} =0\,, \qquad
\tilde{n} \cdot \tilde{n}^* =1\,, \la{012}
\eea
where $\b_i= \a_i$    due to the BPS condition $\D_i=J_i$  (cf. \rf{1.19}, \rf{1.30}). 
For the special choice of $n_i$ in \rf{1.22}  this gives  $  C= \frac{1}{2^{(\b_2+\b_1)/2}} $, 
i.e.   reproduces \rf{1.32}.


\subsection{Agreement with  BPS  3-point  correlator  in free gauge theory}


Since the 3-point function of 1/2 BPS operators is protected,  
~\eqref{1.32} must be the same as  the large charge limit of the corresponding expression 
in free  super Yang-Mills   theory. 
The scalar 1/2 BPS operators in $N=4$  
 supersymmetric gauge  theory can be written in terms of the 6-scalars $\Phi^a$  as 
 (see, e.g., \cite{Osborn})
 \be
{\cal O}_J(\tilde{n})  = {\rm tr} ( \tilde n \cdot \Phi)^J =
 \tilde{n}_{a_1} \dots \tilde{n}_{a_J} {\rm tr} (\Phi^{a_1} \dots
\Phi^{a_J})\ ,
\label{o2}
\ee
where  the complex 6-vector $\tilde n $ satisfies the same constraints as in 
 \rf{012}.
These operators 
%
%
%
have canonically normalized 2-point function.\footnote{There is an additional 
factor of $\frac{1}{\sqrt{J}} (\frac{8 \pi^2}{\lambda})^{J/2}$ 
in the normalization of the operators in~\eqref{o2} which we ignore here 
but such 
 factors will cancel against similar factors in the propagators in computing 
3-point functions up to terms subleading  for large $J_i$.}
%
In order to compute the 3-point function of the operators \rf{o2}
in free  gauge theory we need to contract the fields
in the three   operators.
 Each contraction of the fields in the operators $i$ and $j$ 
will give rise to a factor $(\td n_i \cdot \td n_j)$. 
The  number of contractions among the three operators is as follows~\cite{Seiberg}.
We have to contract 
$\b_3/2$ fields between the first and the second operators, 
$\b_2/2$ fields between the first and the third operators, and
$\b_1/2$ indices between the second and the third operators.  
Ignoring subleading corrections in the limit of large $J_i$ 
we then get the following expression for the 3-point function  coefficient
in \rf{1.21.00} in free SYM theory 
\be 
C_{_{\rm super YM}} = (\tilde{n}_1 \cdot \tilde{n}_2)^{\b_3/2}\ 
(\tilde{n}_1 \cdot \tilde{n}_3)^{\b_2/2}\ 
(\tilde{n}_2 \cdot \tilde{n}_3)^{\b_1/2}\,, 
\label{o4}
\ee
which is indeed  the same as \rf{011} found in the supergravity approach.

\section{Two-point and three-point functions as  correlators of
  vertex operators
in \adss   string theory}


In the rest of this paper we   will 
 consider the semiclassical computation of 3-point functions in 
\adss string theory.
Let us first review some basic points about the  structure of these correlators 
(see also  the discussion in \ci{bt1}). 

\subsection{General  remarks on the structure of correlation functions}

Consider the tree-level 2-point function 
of string vertex operators labelled by 
 points $\vec{a}_1$ and $\vec{a}_2$ 
of the boundary of $AdS_5$ \foot{For simplicity we shall  consider only
scalar  operators  and ignore fermion field dependence.
Both the worldsheet and the target space will be  assumed to be euclidean.}  
\be
G(\vec{a}_1, \vec{a}_2)= \bra \V(\vec{a}_1) \ \V(\vec{a}_2)\ket =
  \frac{1}{\Omega_M} \int {\cal D}{\mathbb X}
\ e^{-  A_0[{\mathbb X}]}\ \V(\vec{a}_1) \ \V(\vec{a}_2)\,.
\label{2.1}
\ee
Here $V(\vec{a}_1)$ and $V(\vec{a}_2)$ are  integrated vertex operators
\be
\V(\vec{a}_{i}) =\int d^2 \xi_{i}\  V\big(z(\xi_{i}), \vec{x} (\xi_{i})-\vec{a}_i; X_p (\xi_i)\big)
\label{2.2}
\ee
where $(z, x_{m})$ are the Poincare coordinates in $AdS_5$ and $X_p$ parametrize $S^5$.
The general structure of $V$
 is 
(ignoring  fermion dependence)
\be
V\big(z(\xi_{i}), x^m (\xi_{i})-a_i^m; X_p (\xi_i)\big)=[K(\vec{a}_i, \xi_{i})]^{\Delta}\
\UU( \xi_{i} )\,,
\label{2.3}
\ee
where $\Delta$ is the target space 
dimension of the operator, $K(\vec{a}_i)$ is  the same as in ~\eqref{1.4},
%
%
and $\UU$ depends on the remaining quantum numbers (spins, etc.).
In~\eqref{2.1}  the integral is over all the \adss string sigma model fields 
with the conformal-gauge  action
\be
A_0[{\mathbb X}]=\frac{\sqrt{\lambda}}{\pi}\int d^2 \xi\ L =
\frac{\sqrt{\lambda}}{\pi} \int d^2 \xi\ \Big( \frac{\partial z \bar \partial z+\partial \vec{x} 
\bar \partial \vec{x}}{z^2}
+L_{S^5} + {\rm fermions}\Big)\,.
\label{2.5}
\ee
As we are considering a tree-level  approximation in closed string theory 
$\xi$ parametrizes a complex plane  and 
 $\Omega_M$ 
\be 
\Omega_M = \int \frac{d^2 \xi_1 d^2 \xi_2 d^2 \xi_3}{|\xi_1-\xi_2|^2|\xi_2-\xi_3|^2|\xi_3-\xi_1|^2}\,,
\label{2.6}
\ee
is the volume of the   $SL(2, {\mathbb C})$ Mobius group, 
which represents the residual gauge transformations
(global conformal diffeomorphisms).   
Assuming that  vertex operators are marginal, i.e. $\D$ is an appropriate function of spins and other
quantum numbers, the worldsheet conformal invariance 
implies that the integral over $\xi_1$ and $\xi_2$ in~\eqref{2.1}  should  factor out as 
\be
\Omega_2 =\int \frac{d^2 \xi_1 d^2 \xi_2}{|\xi_1-\xi_2|^4}\,,
\label{2.7}
\ee
i.e. we should get 
%
\be
G(\vec{a}_1, \vec{a}_2) = \frac{1}{\Omega_c}\ \td G(\vec{a}_1, \vec{a}_2)\  , \ \ \ \ \ \ \ \ \ 
\Omega_{c}\equiv \frac{\Omega_M}{\Omega_2}
\label{2.8}
\ee
%
%
%
where $\Omega_{c}$ represents the volume of the subgroup 
of $SL(2, {\mathbb C})$  that  preserves two points $\xi_1,\xi_2$.  As this subgroup 
is non-compact,  $\O_{c}$  diverges. In flat space 
  this implies the vanishing of the 2-point function.
In the case of string theory in $AdS_{d+1}$, however, the  action  
 has non-compact   global invariance group 
$SO(1,d+1)$. Assuming that vertex operators represent  conformal 
primary fields, the  path integral produces a divergent factor of volume of residual transformations
of $SO(1,d+1)$ that preserve two   fixed boundary points $\vec a_1, \vec a_2$. 
This   factor cancels against the world-sheet factor $\Omega_{c}$ 
producing a finite result  for the  2-point function. 
This point  was discussed in \ci{giv,deb,ku,mo} in the context of string theory 
 in $AdS_3$ based on the corresponding  WZW model  but it applies in general 
 to strings in $AdS_{d+1}$ \ci{bt1}.
 
  We shall explain this in  detail in Appendix A. In particular, 
  in  the context of the semiclassical expansion we are interested in here, 
  the (divergent) volume of residual $SO(1,d+1)$ transformations will appear 
  from an integral over the collective coordinates of a classical solution one 
  is expanding around. 
 

Similarly, in  the case of the  3-point function
\be \la{333}
G(\vec{a}_1, \vec{a}_2, \vec{a}_3 )= \bra\V(\vec{a}_1) \ \V(\vec{a}_2)\ \V(\vec{a}_3)\ket 
\ee
 the integral over 
the operator insertion points $\xi_i$ 
will factorise producing a factor  that will  cancel $\Omega_M$ \rf{2.6} in the denominator.
In the case when   3 target space  points $a_i$ are fixed the remaining symmetry 
subgroup of $SO(1,d+1)$ is  compact $SO(d-1)$ and thus  the resulting correlator is finite.


\subsection{Review of semiclassical computation of  two-point function}


Let us now  review the semiclassical computation of 2-point function of  vertex operators
with large charges 
considering for simplicity the  example of BMN states~\cite{BMN} or chiral primary operators;  
 other examples can be found in~\cite{E1,bt1}.
The  corresponding   vertex operators 
can be written as
%
\bea
&&
\V_{J}(\vec a_1)=\int d^2 \xi_1 \Big[\frac{z}{z^2 +(\vec x-\vec a_1)^2}\Big]^{\Delta}\ (X_1+ i X_2)^{J} \
\VV\,, 
\nonumber \\
&&
\V_{-J}(\vec a_2) =\int d^2 \xi_2 \Big[\frac{z}{z^2 +(\vec x-\vec a_2)^2}\Big]^{\Delta}\ (X_1- i X_2)^{J}\
\VV \,.
\label{3.1}
\eea
Here $V_{-J}\equiv V^*_J$
and $\VV$ stands  for 2-derivative  and fermionic terms that are not relevant 
 for determining the stationary point solution.
The marginality condition implies $\D= J$. 

We shall assume that 
$\vec a_i = (a_i, 0, 0, 0)$; then, jumping ahead, one can argue  that 
 the semiclassical trajectory will 
belong to $AdS_2$, i.e. we can set $\vec x = (x, 0, 0,0)$. 
Also, if we set  (as in \rf{1.14})  $X_1+ i X_2= \cos \psi \ e^{i \varphi}$, 
then on the semiclassical trajectory $\psi=0$, i.e. 
we may replace $(X_1\pm  i X_2)^J$ by $e^{\pm i J\varphi}$.

In the limit of large $\Delta, J$, the 2-point function
%
%
is governed by the semiclassical trajectory with singularities prescribed by the vertex operators.
It can be found  from the  ``effective'' action\foot{We use the notation 
$\partial = {1 \ov 2} ( \partial_1 - i  \partial_2), \ 
\bar \partial = {1 \ov 2} ( \partial_1 + i  \partial_2)$.}
\bea
A& = &A_0 - \ln V_{J}(a_1) - \ln V_{-J}(a_2)
\nonumber \\ 
& = & \frac{\sqrt{\lambda}}{\pi} \int d^2 \xi\ \Big[ \frac{1}{z^2}(\partial z \bar \partial z+
\partial x \bar \partial x) + 
\partial \varphi \bar \partial \varphi\Big]
\nonumber \\
&  &- \D \int d^2 \xi\ \Big[\delta^2(\xi-\xi_1)\ \ln \frac{z}{z^2+(x-a_1)^2} +
 \delta^2(\xi-\xi_2)\ \ln \frac{z}{z^2+(x-a_2)^2} \Big]
\la{3.44}\\
& &- i J \int d^2 \xi\ \big[ \delta^2 (\xi-\xi_1) -\delta^2 (\xi-\xi_2)\big]\ \varphi\, +... \ ,
\label{3.4}
\eea
where  $A_0$ is the classical string action 
 \rf{2.5} where we set to zero all irrelevant fields. Dots stand for  
  $\ln \VV$  terms subleading at large $\D=J$. 
 The semiclassical expression  for the 2-point  function can then be written as
 (see also Appendix A;  here we assume that collective coordinate contribution is  absorbed into $\GG$)
 \be \la{3.2}
 G(\vec a_1,\vec a_2)  = { 1 \ov \Omega_M} 
 \int d^2 \xi_1 d^2 \xi_2 \ \GG(\vec a_i; \xi_i) \ , \ \ \ \ \ \ \
 \GG\ \sim \   e^{- A }  \ .
 \ee
To find the stationary point trajectory \ci{pol,t,E1,bt1}
we  may  start with 
the euclidean 
version of the corresponding classical solution on the cylinder $(\tau, \sigma)$
which carries the same charges
as the vertex operators and  then  transform this solution to the complex  $\xi$-plane 
by the  conformal map 
\be 
e^{\tau +i \s} =\frac{\xi-\xi_1}{\xi-\xi_2}\,. 
\label{3.5}
\ee
%
 In this construction   all the information 
about the singularities at $\xi_i$  is encoded in the conformal map \rf{3.5}. 

In the present case of the BMN states the relevant classical solution is the  (analytically continued) 
 geodesic
connecting the points $x=a_1$ and $x=a_2$ in $AdS_2$ (for concreteness we shall assume $a_2 >a_1$)
\bea
&&
z=\frac{a_2-a_1}{2 \cosh (\kappa \t)}\,, \qquad 
x =\frac{a_2-a_1}{2} \tanh (\kappa \tau) +\frac{a_2+a_1}{2} \,, \qquad \kappa =\frac{\D}{\sqrt{\lambda}}\,, 
 \label{3.6}\\
&&
\varphi=-i \omega \tau \,, \qquad \qquad \omega=\frac{J}{\sqrt{\lambda}}\,.
\label{3.66}
\eea
%
One can explicitly check that  \eqref{3.5},\eqref{3.6},\rf{3.66} indeed solve the equations 
following from    \rf{3.4}.
 The equation for $\varphi$ reads
\be
\bar \pt \pt \varphi =
\frac{i J \pi}{2\sqrt{\lambda}} \ \big[\delta^2 (\xi-\xi_2) -\delta^2 (\xi-\xi_1)\big]\,.
\label{3.7}
\ee
The solution to this equation is 
\be
i \varphi = \frac{J}{ \sqrt{\lambda}}\ \big( \ln |\xi-\xi_1| - \ln |\xi-\xi_2|\big)
\label{3.8}
\ee
which is precisely $\omega \tau$ if we use the map~\eqref{3.5}. 

Let us point out a  subtlety which will be important later
when we consider 3-point functions. 
 In two dimensions solutions to the Laplace equation with 
  prescribed singularities  like \rf{3.7},   
in general, do not go 
to zero at infinity. 
That means  there  may be 
 an additional unwanted singularity at $\xi=\infty$. 
Indeed, if the charges of the vertex operators 
in~\eqref{3.1} were different, $J_1 \neq J_2$,  instead of \rf{3.7}
we would have
%
\be
\bar \pt \pt \varphi =
\frac{i  \pi}{2\sqrt{\lambda}} \ \big[J_1 \delta^2 (\xi-\xi_2) -J_2\delta^2 (\xi-\xi_1)\big]
\label{3.8.1}
\ee
with the solution being 
%
\be
i \varphi = \frac{1}{ \sqrt{\lambda}}\ \big( 
J_1\ln |\xi-\xi_1| - J_2\ln |\xi-\xi_2|\big)\,.
\label{3.8.2}
\ee
%
Then  $\varphi$ would have a logarithmic singularity not only at 
$\xi=\xi_1, \xi_2$ but also at $\xi=\infty$. 
One  interpretation of this  could be that 
we have an additional vertex operator inserted at   $\xi=\infty$  
whose charge is $J_2-J_1$
so that the total charge remains zero.
The condition that the solution is 
non-singular at infinity (i.e. is properly defined on a   2-sphere)
is precisely $J_1=J_2$.
It can be  derived by integrating  both sides of \rf{3.8.1}
over the complex  plane (i.e. 2-sphere that  has no boundary).  
A heuristic way to 
  arrive at same   condition is 
by looking at the right hand side of~\eqref{3.8.1} and demanding that 
it does not have a delta-function
source at  large $\xi$:
for that  one  can ignore $\xi_1$ and $\xi_2$ compared to $\xi$ in
the delta-functions in~\eqref{3.8.1} and require 
 that the coefficient in front of the 
resulting $\d^2(\xi)$ (namely,  $J_1-J_2$) is zero.

The equations for $x$ and $z$ are substantially more complicated 
and without knowing the relation to the classical solution~\eqref{3.6},\rf{3.66}  it would seem 
 hard to solve them. The equation for $x$ is 
\be 
\pt \big(\frac{\bar \pt x}{z^2}\big) +
\bar \pt \big(\frac{ \pt x}{z^2}\big) =\frac{ 2\pi \D}{\sqrt{\lambda}} \Big[
\frac{x-a_1}{z^2 +(x-a_1)^2} \d^2 (\xi-\xi_1) 
+ \frac{x-a_2}{z^2 +(x-a_2)^2} \d^2 (\xi-\xi_2)\Big] \,.
\label{3.9}
\ee
When we substitute~\eqref{3.6},\rf{3.66},\eqref{3.5} into~\eqref{3.9} we find that both sides of it  become
equal to 
\be
\frac{ 2\pi \D}{(a_2-a_1)\sqrt{\lambda}} \ \big[\d^2 (\xi-\xi_1) -\d^2 (\xi-\xi_2)\big]\,.
\label{3.10}
\ee
The equation for $z$ can also be shown to be satisfied in a similar way 
(see \cite{bt1} for details). 
As before, let us point out that eq.~\eqref{3.10} is non-singular when 
$\xi \to \infty$ meaning that our solution does not have an unwanted singularity 
at infinity. This is achieved because 
\be
\frac{x-a_1}{z^2 +(x-a_1)^2}\Big|_{_{\xi\to \xi_1}} =- 
\frac{x-a_2}{z^2 +(x-a_2)^2}\Big|_{_{\xi\to \xi_2}}
\label{3.10.1}
\ee
for the geodesic~\eqref{3.6}. Let us also note that these combinations 
are constants along~\eqref{3.6} so~\eqref{3.10.1} turns out to be satisfied
for any $\xi$. 

Evaluating the action ~\eqref{3.4} on this solution we get 
\be
 e^{-A}\ \sim\ \frac{1}{(a_2-a_1)^{2 \D}} \ 
 |\xi_1-\xi_2|^{\frac{\Delta^2-J^2}{\sqrt{\lambda}}}\,.
\label{3.11}
\ee
In computing the action we subtracted the 
divergences of the form  $\ln |\xi-\xi_i|$ with  $\xi \to \xi_i$
corresponding to self-contractions in the vertex operators.\foot{They should automatically go away if the 
vertex operators are defined  with an appropriate ``normal ordering'', i.e. as
proper marginal operators.}
In addition, $\GG$ in \rf{3.2} 
contains  a  factor  of $|\xi_1-\xi_2|^{-4}$ coming from the expectation 
value of the 2-derivative
terms  $\VV$ in \rf{3.1}. As a result,
taking into account the marginality condition $\D=J$   we recover the 
factor  $\Omega_2$ \rf{2.7} as required by 2d conformal invariance;  
it  cancels out as explained in  the previous subsection and Appendix A. 
Thus  for $\D = J \sim \sql \gg 1$ we  finish with 
\be
G(\vec a_1,\vec a_2)  =\frac{1}{|\vec a_1-\vec a_2|^{2 \D}}
\label{3.13}
\ee
up to possible subleading corrections  
depending on proper normalization of the vertex operators.

Let us note that while written  in the $\xi$-coordinates the solution \eqref{3.6} looks  rather complicated, 
 the map~\eqref{3.5} ``trivializes'' it. 
This point will be important in the subsequent discussion of the semiclassical 
3-point functions.

\subsection{Correlator of 3 chiral primary  operators at strong coupling}

The vertex operator  of a chiral primary state is parametrized 
by a point $\vec a$ at the AdS boundary and a complex null 6-vector $n_p$.
Instead  of $\vec a$ to label the boundary point 
 we may use  a null 6-vector $\nN_r$ as discussed in 
section 2.4 (see \rf{02}).\foot{Similar ``embedding''  parametrization
 is often useful  in 4d CFT
to make the action of the conformal group
 $SO(2,4)$ linear (see, e.g., \ci{co}
and references there).}
Then the general form of the string vertex operator  \rf{3.1} 
representing the highest weight $[0,J,0]$ gauge theory chiral primary operator \rf{o2}, i.e.  
${\cal O}_J (\nN; n) = \tr  [n \cdot \Phi(\nN)]^J$,  is\foot{In our  notation  
(see \rf{01})\ 
 $(\nN, Y)= - \nN^r Y_r\ , \ \ Y^r Y_r
 = - Y_{-1}^2 + Y_m Y^m - Y_{4}^2=-1, \ \    \nN^r \nN_r=0$. Here we  use $n_i$ 
 instead of $\tilde n_i$ in \rf{012},\rf{o2}.
  }
\bea
\V_{J}(\nN;n)=
\int d^2 \xi \  (\nN, Y)^{-\Delta}\ (n \cdot X)^J  \
\VV\,, \ \ \ \ \ \ \ \ \ \ \ \ \ \ \D= J \ . 
\label{3.14}
\eea
 $\VV$ represents again appropriate  2-derivative  and fermionic terms 
 that ensure  marginality  for $\D= J$  (and proper normalization). 
 
Suppose we would like to compute the  correlator like  \rf{333}, i.e. 
$G= 
\bra\V_{J_1} (\nN_1;n_1) ...\V_{J_k} (\nN_k;n_k)   \ket 
$
in \adss string theory   with the action \rf{3.4}  written   in the 
embedding coordinates  as follows:
$\frac{\sqrt{\lambda}}{\pi} \int d^2 \xi\ \Big[
 \partial Y^r \bar \partial Y_r+  \Lambda ( Y^r Y_r+1)  
+\partial X_p \bar \partial X_p
+  \tilde \Lambda ( X_p X_p -1)  + {\rm fermions}\Big]. $
If we split $Y_r$ and $X_p$  as well as $\Lambda$ and $\td \Lambda$ 
into the constant 0-mode parts 
and the non-constant  fluctuation parts, then the  contribution of the latter parts 
will be   suppressed in the  strict large tension limit (more precisely, they 
will produce just overall  determinant factors).\foot{We are grateful to S. Giombi for a discussion of this 
point.}
Similarly, the fermionic couplings 
and the the contractions  involving the  
 dimension (1,1)  factors $\VV$ in \rf{3.14}  will also be subleading. 
Then   the correlator will  formally factorizes
 into the one involving the 0-mode  $Y_{r0}$ and the one involving the 0-mode   $X_{p0}$.
The two   integrals   are  in general    related by the analytic continuation
 $ Y_m \to i X_m , \ 
\sqrt{\lambda} \to - \sqrt{\lambda}$, etc. 
 The  free-theory correlator of the factors $(n_1 \cdot X(\xi_1))^{J_1} ...
 (n_k \cdot X(\xi_k))^{J_k} $  is readily computed if  all $J_i$ are integer.
 For example, in the 3-point case we get the same factor as in \rf{o4}, 
 i.e.\foot{This is essentially the same 
 computation as in free gauge theory mentioned in section 2.5 
 where instead of the six  2d scalars $X_p$ we have the six 4d scalar 
fields  $\Phi_p$ which are matrices in adjoint representation 
of $SU(N)$.}
$({n}_1 \cdot {n}_2)^{\a_3/2}\ 
({n}_1 \cdot {n}_3)^{\a_2/2}\ 
({n}_2 \cdot {n}_3)^{\a_1/2}$, where $\a_1= J_2 + J_3 - J_1 $, etc. 
 To evaluate  the correlator of  $(\nN_1, Y(\xi_1))^{-\Delta_1} ...
 (\nN_k, Y(\xi_k))^{-\Delta_k}   $
 one can formally  continue $\D_i$ to negative integer  values and then  continue 
 back.
 The remaining integrals over the 0-modes of $ \Lambda$ and $\tilde \Lambda$ 
 will be very similar  and will essentually cancel each other because of the 
  marginality  condition  $\D_i=J_i$.  
  We will then  end up with 
\be 
G_{_{\sql \gg 1}}  \sim  \Big[ {
({n}_1 \cdot {n}_2) \ov (\nN_1, \nN_2)}\Big]^{\a_3/2}
 \Big[{({n}_1 \cdot {n}_3) \ov (\nN_1, \nN_3)}\Big]^{\a_2/2}
 \Big[{({n}_2 \cdot {n}_3) \ov (\nN_2, \nN_3)}\Big]^{\a_1/2}\ . \la{313}
 \ee
This  remarkably symmetric form of the 3-point correlator 
 is  the same  as \rf{1.21.00}   with \rf{o4},\foot{Note that 
  $(\nN_1, \nN_2)= 2 (\vec a_1 - \vec a_2)^2$, etc., 
   cf. \rf{02},\rf{05},\rf{09}.} 
which is, of course,  not surprising 
 as the  correlator  of 3 CPO's should not be
renormalized, i.e. should be the same at weak and at strong coupling. 
In fact, the contributions of $\VV$ factors and fermions should 
conspire so that the  $\sql \gg 1$ 
string result ``localises'', i.e. reduces to the one in 
the  supergravity approximation.

Below   we shall demonstrate how to 
reproduce the  same  result \rf{313} in the case when $J_i$ are as large as string 
tension\foot{Note that in this limit the supergravity approximation may, in general,
fail;
this does not happen, of course, 
 in the case of protected 3-point function of CPO's.}
using    semiclassical approximation in string theory path integral.


\section{Semiclassical computation of extremal 3-point function}


We shall study a  semiclassical computation  of   the 3-point functions 
with the extremal  case when $\D_1=\D_2+\D_3$.
Here we shall  explicitly consider  the correlator of BPS states 
but the general discussion   of the $AdS_5$  contribution 
given below would formally apply also to the case 
of non-BPS operators with non-trivial charges  in $S^5$  and  having $\D_1=\D_2+\D_3$.

In the extremal case we may assume that all  three BPS operators 
carry charges in the same $SO(2)$   subgroup of $SO(6)$ symmetry of $S^5$.
Starting  with the  operators like in \rf{3.1}  with $\D_i = | J_i|$ 
and  $\vec a_i=(a_i, 0, 0, 0)$\foot{This choice is  always allowed as 
the general dependence of the  correlator \rf{333}  on $\vec
a_i$ is fixed by conformal  invariance to be as in \rf{1.21.00}.}
and being interested only in the leading semiclassical contribution 
we may choose  them in the form 
\be
\V_{J_i}(\vec a_i)=\int d^2 \xi\ \Big[\frac{z}{z^2 +(x-  a_i)^2}\Big]^{\Delta_i}\  e^{i J_i \varphi} \ 
\VV_i\,, \la{41}
\ee
where we set to 0 all ``irrelevant'' coordinates that vanish on the 
semiclassical trajectory.
 We shall also choose $a_i$ as 
%
%
$a_1=0<a_2<a_3$. The integral over the zero mode of $\varphi$ then imposes
charge conservation, i.e.  we may consider 
%
\be
G(a_1, a_2, a_3)=\langle V_{J_1}(a_1) V_{-J_2}(a_2)V_{-J_3} (a_3)\rangle \ , \ \ \ \ \ \ \ \ 
J_1=J_2+J_3\,, \qquad \D_i=J_i\,.
\label{4.2}
\ee
%
%
%
In the semiclassical limit ($J_i \sim \sql \gg 1$) 
of the correlation function~\eqref{4.2} 
is controlled by the  extremum  of 
the  following    action  (cf. \rf{3.44}) 
\bea
&& A=A_{AdS}+ A_{S^5}\,,\nonumber \\
&& A_{AdS}=  \frac{\sqrt{\lambda}}{\pi} \int d^2 \xi\ \frac{1}{z^2}(\partial z \bar \partial z+
\partial x \bar \partial x)
\nonumber \\
&& \ \ \ 
-\ \int d^2 \xi\ \Big[  \D_1  \delta^2(\xi-\xi_1) \ln \frac{z}{z^2+(x-a_1)^2}
+ \D_2 \delta^2(\xi-\xi_2) \ln \frac{z}{z^2+(x-a_2)^2}
\nonumber\\
&&\ \ \  \ \ \ \ \ \  \ \ \  \ \ \ \ \ \ +\ \D_3  \delta^2(\xi-\xi_3) \ln \frac{z}{z^2+(x-a_3)^2}\Big]\,, \\
\label{4.4}
&&
A_{S^5} =  \frac{\sqrt{\lambda}}{\pi} \int d^2 \xi\  
\partial \varphi \bar \partial \varphi - i \int d^2 \xi  \Big[ 
J_1 \delta^2 (\xi-\xi_1)  - J_2  \delta^2 (\xi-\xi_2)
- J_3  \delta^2 (\xi-\xi_3)\Big] \ \varphi  \,. \nonumber 
\eea
We shall first find  the solution in $S^5$ and then consider the  $AdS_5$ part.


\subsection{Solution in $S^5$}


The equation of motion for the angle $\varphi$ 
\be
\partial \bar \partial \varphi 
=-\frac{i \pi}{2\sqrt{\lambda}} \big[ J_1 \delta^2 (\xi- \xi_1) -J_2 \delta^2 (\xi- \xi_2)
-J_3 \delta^2 (\xi- \xi_3)\big]\,
\label{4.5}
\ee
is solved by 
\be
\varphi= -{i\ov \sql} \Big(J_1 \ln |\xi-\xi_1| 
-J_2\ln |\xi-\xi_2|  -J_3\ln |\xi-\xi_3|\Big) \,.
\label{4.6}
\ee
Like in the case of the 2-point function (cf. \rf{3.6}--\rf{3.8}) 
let us  introduce a new coordinate $\tau$ such that\foot{In general, one may  start with 
 $\varphi = -{i\ov \sql} J_1 (\tau  -\hat{\tau})$  
 but the constant $\hat{\tau}$ can be absorbed into the   shift of the origin of $\tau$
 or constant shift of $\varphi$.  In what follows  we shall  set $
 \hat{\tau}=0$  to simplify the formulae.}
%
\be
 \varphi = -i \omega_1  \tau \ , \ \ \ \ \ \ \ \ \ \ \ \ \omega_1 = {J_1\ov \sql} \ ,  
\label{4.8}
\ee
i.e.  define the following map from the complex plane $\xi$ with three marked points 
to a complex domain $(\tau,\s)$ 
\be
\zeta =e^{\tau + i \sigma} 
=\frac{\xi-\xi_1}{(\xi-\xi_2)^{J_2/J_1} (\xi-\xi_3)^{J_3/J_1}}\,.
\label{4.9}
\ee
%
Here the points $\xi_1, \xi_2, \xi_3$ are mapped to either $\tau=-\infty$ 
or $\tau=+\infty$. Note that since $J_1=J_2+ J_3$ we do not have an additional 
singularity at $\xi=\infty$. 
This is, in fact,  a familiar Schwarz--Christoffel
map
from  a  plane with 3 punctures 
into the ``light-cone''  three  closed strings interacting diagram in flat space \ci{man}
(with  one cylinder at $\tau=-\infty$  becoming  two joined cylinders at $\tau=\infty$).
Here  
the role of conserved components of
the light-cone  momenta $p^+_i$ or lengths of the three strings in the
light-cone gauge  is played by $J_i$, i.e. by the components of the 
  angular momentum along $S^1 \subset S^5$.

To simplify the discussion we may first  replace the  cylinders  by strips  by 
   cutting  each  cylinder along the $\tau$-direction
and view it as two copies of an infinite  strip 
(imposing   periodicity on functions of $\s$  at the end). 
For example,  an infinite strip of width $\pi$ is mapped by~\eqref{3.5} 
 to the upper half plane with two marked points 
$\xi_1, \ \xi_2$  lying on the  the real axis. 
In general, conformal transformations from the upper half plane with marked points  
to the interior of a polygon 
are known as Schwarz--Christoffel maps  with 
~\eqref{4.9} being  a simple example. 
Let us review  how the complex domain parametrized by $(\tau, \sigma)$ 
can be found in the case of ~\eqref{4.9}.
Let us assume  for concreteness that   the points $\xi_i$ on the real axis 
are ordered as  $\xi_1<\xi_3<\xi_2$  
and  start moving from  $\xi >\xi_2$ in the direction of decreasing $\xi$.
Once we cross $\xi_2$ and start moving towards $\xi_3$ we pick up a phase 
$e^{i \pi J_2/J_1}$ meaning that $\sigma$ has jumped by $\pi J_2/J_1$. 
This means that we cannot reach $\xi_3$ unless $\sigma > \pi J_2/J_1$.
This, in turn, means that we have a cut along along the $\tau$-direction starting at 
the some point $(\tau_{int}, \sigma_{int})$ with $\sigma_{int}= \pi J_2/J_1$. 
The points $\xi_2$ and $\xi_3$ lie on the opposite sides of the cut (see Figure 2). 
\begin{figure}[ht]
\centering
\includegraphics[width=70mm]{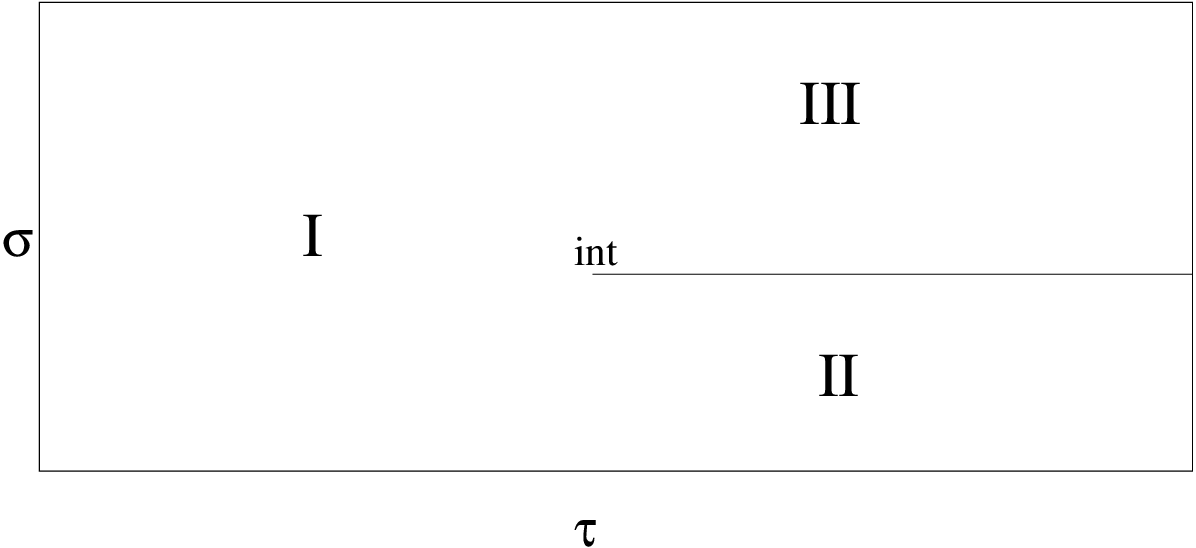}
\caption{\small The 
$(\tau, \s)$  domain which is mapped
to the upper half plane with three marked points by  the Schwarz-Christoffel map~\eqref{4.9}.
The regions I, II, III can be identified with the three interacting strings. 
The length of the strings is set by the angular momenta $J_i$
that  satisfy  $J_1=J_2+J_3$.
}
\label{fig2}
\end{figure}
The point $(\tau_{int}, \sigma_{int})$ may be 
 interpreted as the interaction point,
where one incoming string splits into the two outgoing strings. 
It  can be found  as the critical point of the map~\eqref{4.9}:
\be
\frac{\pt \zeta }{\pt \xi}=0\,.
\label{4.10}
\ee
Using  $J_1=J_2+J_3$ we obtain a linear equation for $\xi=\xi_{int}$ 
solved by 
\be
\xi_{int}= \frac{J_2 (\xi_1-\xi_2) \xi_3 + J_3 (\xi_1-\xi_3) \xi_2}
{J_2 (\xi_1-\xi_2) + J_3 (\xi_1-\xi_3) }\,.
\label{4.11}
\ee
Substituting it back  into~\eqref{4.9} we get 
\be 
\tau_{int} + i \sigma_{int} 
=\ln  \frac{(\xi_1 - \xi_2)^{J_3/J_1}(\xi_3 - \xi_1)^{J_2/J_1}}{(\xi_3 - \xi_2)}
+ \ln \frac{J_1}{J_2^{J_2/J_1}J_3^{J_3/J_1}}
\,,
\label{4.12}
\ee
so that for the above choice of $\xi_i$  we have 
 $\sigma_{int}= \pi J_2/J_1$.
 Note that the value of $\tau_{int}$ is  unphysical and one can shift it, e.g., to  zero  
by re-introducing a constant  shift of $\tau$ in~\eqref{4.9}.  

The  discussion in the previous paragraph and  Figure 2 applied to open strings
and has an advantage that it is easier to visualize. 
 The case of  closed strings can be described by  doubling trick,
 to take two copies of the domain in Figure 2
and perform appropriate identifications 
to ensure periodicity in  $\s$. 
The resulting  domain will  be mapped to the complex plane with three marked points using~\eqref{4.9}.

Explicitly, the  3 regions  of the  $(\tau, \sigma)$ domain  in Figure 2 
representing 3 interacting strings  are 
\bea
&&
I:\ \ \ \   \tau \in (-\infty, 0]\,, \quad \s \in [0, \pi]\,,  
\nonumber \\ &&
II:\ \ \ \   \tau \in [0, +\infty)\,, \quad \s \in [0, \sigma_{int}]\,, 
\nonumber \\ &&
III:\ \ \ \    \tau \in [0, +\infty)\,, \quad \s \in [\sigma_{int}, \pi]\,.
\label{4.13}
\eea
Doubling the $\s$-intervals 
we can find the angular momenta  of  the corresponding   closed strings as 
\bea
&&
J_1=2 i \frac{\sqrt{\lambda}}{2 \pi} \int_0^{\pi} d \sigma \ \pt_{\tau} \varphi \ , 
 \ \ \  \ \ \ \ \ 
J_2=2 i \frac{\sqrt{\lambda}}{2 \pi} \int_0^{\s_{int}} d \sigma \ \pt_{\tau} \varphi =
J_1 \frac{\s_{int}}{\pi}\,, \ \ \
  \nonumber \\ && \ \ \ \  \ \ \ \  \ \ \ \ 
J_3=2 i \frac{\sqrt{\lambda}}{2 \pi} \int_{\s_{int}}^{\pi} d \sigma \ \pt_{\tau} \varphi =
J_1  \frac{\pi- \s_{int}}{\pi}\,.
\label{4.14}
\eea
Here $\frac{\sqrt{\lambda}}{2 \pi}$ is the string  tension and factor of 2 is due to the doubling 
of the $\s$ interval. 
 We thus  have again 
 $\sigma_{int}= \pi J_2/J_1$.

Finally, computing 
the $S^5$ part of the action in \rf{4.4}  on the solution~\eqref{4.6} we find 
\be
A_{S^5}(\xi_1,\xi_2,\xi_3) = \frac{1}{\sqrt{\lambda}}\Big(J_1 J_2 \ln |\xi_1-\xi_2| +
J_1 J_3 \ln |\xi_1-\xi_3| -J_2 J_3 \ln |\xi_2-\xi_3|\Big)\,,
\label{4.15}
\ee
where we omitted  logarithmic ``self-contraction''  divergences $\ln |\xi-\xi_i|_{\xi \to \xi_i}$.


\subsection{Solution in $AdS_5$}

Let us now  consider the solution of the  equations of motion for $z$ and $x$ following from 
\rf{4.4}:
\bea
&&
\partial \big( \frac{\bar \partial x}{z^2}\big) + \bar \partial \big(
 \frac{\partial x}{z^2}\big)=\frac{2 \pi \D_1}{\sqrt{\lambda}}\Big[
\D_1  \frac{x}{z^2+x^2} \delta^2 (\xi-\xi_1) 
\nonumber \\
&& \ \ \ 
+\ \D_2 \frac{x-a_2}{z^2+(x-a_2)^2} 
\delta^2 (\xi-\xi_2) 
+ \D_3 \frac{x-a_3}{z^2+(x-a_3)^2} 
\delta^2 (\xi-\xi_3)\Big]\,,
\label{4.17}\\
&&
\pt \big( \frac{\bar \pt z}{z^2} \big)
+\bar \pt \big( \frac{\pt z}{z^2} \big)
 +\frac{2}{z^3} (\pt z \bar \pt z + \pt x \bar \pt x)= 
\frac{\pi }{\sqrt{\lambda}} \Big[ \D_1 \frac{z^2- x^2}{z^2+ x^2} \d^2 (\xi-\xi_1)
\nonumber \\
&&\ \ \ 
+\ \D_2 \frac{z^2 -(x-a_2)^2}{z^2+ 
(x-a_2)^2} \d^2(\xi-\xi_2)
+ \D_3 \frac{z^2 -(x-a_3)^2}{z^2+ (x-a_3)^2} \d^2 (\xi-\xi_3)\Big]\,.
\label{4.18}
\eea
As was discussed in the previous section below eqs.~\eqref{3.8} and~\eqref{3.10}, 
the solution to these equations,
in addition 
to the singularities at $\xi_1$, $\xi_2$, $\xi_3$, might also have a 
singularity at $\xi=\infty$.
We can demand its absence by studying
 how the right-hand-sides of~\eqref{4.14},\eqref{4.15}
behave at large $\xi$.\footnote{If in the formal limit of large $\xi$ the
 right-hand-sides of~\eqref{4.14},\eqref{4.15}
remain singular the solution is expected to be  singular at $\xi=\infty$. 
This may be effectively attributed to  the presence of an additional 
vertex  operator at infinity.}
 This  suggests that one  should impose 
the 
two equations analogous to eq.~\eqref{3.10.1}
%
\bea
&&
\D_1\frac{ x}{z^2+ x^2}\Big|_{_{\xi\to\xi_1}} + \D_2 \frac{x-a_2}{z^2+ (x-a_2)^2}
\Big|_{_{\xi \to \xi_2}}
+  \D_3 \frac{x-a_3}{z^2+ (x-a_3)^2}\Big|_{_{\xi \to \xi_3}} =0\,, \nonumber \\
&&
\D_1 \frac{z^2- x^2}{z^2+ x^2}\Big|_{_{\xi \to \xi_1}} + \D_2 \frac{z^2 -
(x-a_2)^2}{z^2+ (x-a_2)^2}\Big|_{_{\xi\to \xi_2}}
+ \D_3 \frac{z^2 -(x-a_3)^2}{z^2+ (x-a_3)^2}\Big|_{_{\xi \to \xi_3}}
=0
\label{4.19}
\eea
These equations will be indeed satisfied on the solution we are going to construct.


Let us  now 
 show that the solution to eqs.~\eqref{4.17}, \eqref{4.18}, \eqref{4.19}
can be obtained by combining the
conformal map \rf{4.9} from the complex plane with 3 marked points to the  3-cylinder 
double of Figure 2
with  the  construction  of intersection of 3 geodesics in $AdS_2$ in 
\cite{km}. See Figure 3. 
\begin{figure}[ht]
\centering
\includegraphics[width=90mm]{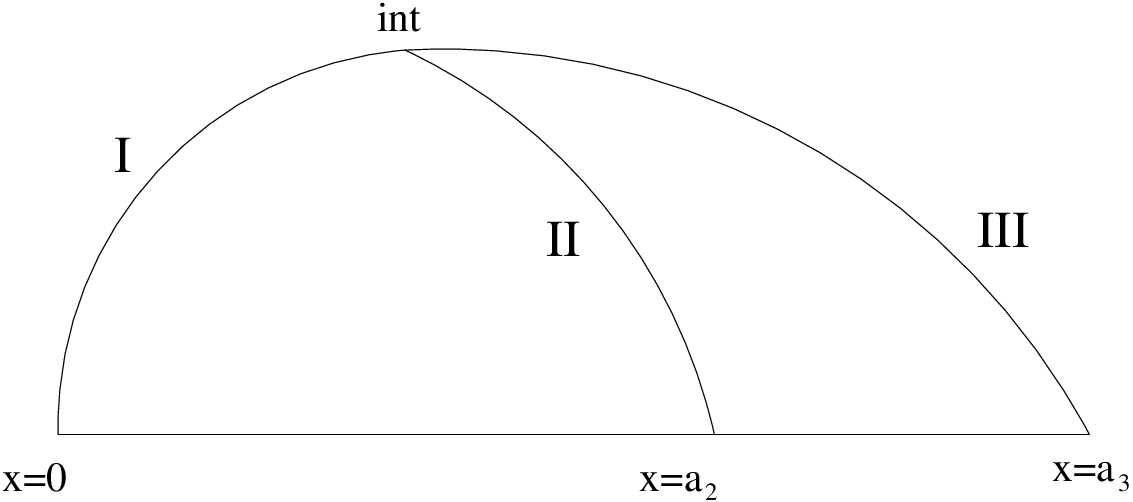}
\caption{\small Three geodesics in $AdS_2$ meeting at the interaction point.}
\label{fig3}
\end{figure}
The $\tau$-parameter  of the  three intersecting geodesics will be related to 
$\xi_i$ by  a map similar to \rf{4.9} 
%
\be
\zeta =e^{\tau+ i \sigma} =\frac{\xi-\xi_1}{(\xi-\xi_2)^{\D_2/\D_1} (\xi-\xi_3)^{\D_3/\D_1}}\,.
\label{4.16}
\ee
%
Note that this map is well-defined (no additional singularity at $\xi=\infty$) 
only if $\D_1=\D_2+\D_3$.\foot{In the non-extremal case  we will have
 to use  a different Schwarz--Christoffel map discussed in the next section.}
For BPS states  \rf{4.16} is actually equivalent to \rf{4.9} due to the marginality conditions $\D_i=J_i$. 

We will construct the full solution everywhere in the domain 
in Figure~2  following the idea  
of \cite{km}, i.e. 
we will define independent solutions in the regions I, II, III 
in \eqref{4.13} and ``glue'' them together  at the interaction point that will correspond
$(\tau,\sigma)=(\tau_{int},\s_{int})$. 
Near each singularity the solution has to approach  a geodesic
of the type~\eqref{3.6}; in the BPS case (and more generally, for a string state 
that does not carry $AdS_5$ charges except energy)
it is natural to propose that 
the solution in each region should, in fact, 
be a piece of a geodesic with  appropriate target space boundary conditions. 

First, let us make sure that the three intersecting 
geodesics are compatible with eqs.~\eqref{4.19}. 
This compatibility follows from the fact, discussed in the previous section, 
that each term in eqs.~\eqref{4.19}
is a constant along the geodesic that 
originated at $a_i$ (i.e. corresponding to  $\xi_i$). 
Thus we can evaluate all the 
terms in~\eqref{4.19} at the same point $\xi_{int}$. But 
then these equations 
can be viewed as the conditions for the intersection point in the 
target space
 $(z_{int}= z(\xi_{int}), x_{int}=x(\xi_{int}))$.
With this interpretation, 
these  are  the same equations as the ones in \rf{1.17} that  extremize
the ``action'' \rf{1.16} appearing in  the supergravity integral in 
section 2.2. 
The solution to these equations is given in~\eqref{1.18}.
Like in section 2 we will have to  assume that  $\Delta_1=\D_2 +\D_3+ \epsilon$, i.e. to go 
slightly off  extremality to lift  the ``interaction'' 
point~\eqref{1.18} from  the boundary and take $\epsilon \to 0$ in the final expressions. 
Note that eqs.~\eqref{4.19} are not the same as~\eqref{1.17}. The former are the functional equations 
rather than algebraic.  However, they reduce to the 
algebraic equations on our geodesic 
ansatz.

Explicitly,  the  solutions 
in the regions I, II, III are expected to be 
\be
I:\  z^2 =x (b_1-x)\,,\qquad II: \ z^2 =(a_2-x) (x-b_2)\,, \qquad
III: \ z^2 =(a_3-x) (x-b_3)
\label{4.20}
\ee
Each geodesic is a half-circle in $(z,x)$ plane  connecting 
 one of the boundary  points $a_i$ with some other boundary points $b_i$.
The values of $b_i$
\bea
b_1 =\frac{(\a_2 +\a_3) a_2 a_3}{\a_2 a_2 + \a_3 a_3}\,, \ \ \ \ \ \ \ \ 
b_2 =\frac{\a_1 a_2 a_3}{(\a_1 +\a_3) a_2 - \a_3 a_3}\,, \ \ \ \ \ \ \ \ 
b_3 =\frac{\a_1 a_2 a_3}{(\a_1 +\a_2) a_3 - \a_2 a_2}\,
\label{4.21}
\eea
 can be found \ci{km} by  demanding that these three geodesics 
meet at the point ($(x_{int},z_{int})$ given in \eqref{1.18}  (see Figure~3).
%
%
Parametrizing each geodesic   by $\tau$ as in \eqref{1.11},\eqref{3.6}
we can thus  write the proposed solution in the $(\tau,\sigma)$ domain
explicitly  as 
\bea
&&
I_{\tau \in (-\infty, \t_{int}]\,, \  \s \in [0, \pi]}  : \qquad \ \ z= \frac{b_1}{2 \cosh( \kappa_1 \tau +\t_1)}\,, \quad 
x= \frac{b_1}{2} \tanh( \kappa_1 \tau +\t_1)+\frac{b_1}{2}\,,
\label{4.22} \\
&&
II_{ \tau \in [\t_{int}, +\infty)\,, \  \s \in [0, \sigma_{int}]    }: \quad\  z= \frac{a_2 - b_2}{2 \cosh( \kappa_2 \tau +\t_2)}\,, \quad 
x= \frac{a_2-b_2}{2} \tanh( \kappa_2 \tau +\t_2)+\frac{a_2+ b_2}{2}\,,
\nonumber \\
&&
III_{ \tau \in [\t_{int}, +\infty)\,, \  \s \in [\s_{int},  \pi]  } : \quad z= \frac{a_3 - b_3}{2 \cosh( \kappa_3 \tau +\t_3)}\,, \quad 
x= \frac{a_3-b_3}{2} \tanh( \kappa_3 \tau +\t_3)+\frac{a_3+ b_3}{2}\,.
\no \eea
%
The parameters $\k_i$ are to be fixed by matching against the singularities
prescribed by the vertex operators. The parameters $\t_i$ are introduced to make sure 
that the three segments of the solution  
intersect at the interaction point $\t=\t_{int}$ which we can always choose to be at zero. 
Demanding that these three geodesics  meet at~\eqref{1.18} for  $\tau=\t_{int}=0$ gives
\bea
&&
\t_1= \frac{1}{2}
\ln \frac{\a_1 (\a_2 a_2 + \a_3 a_3)^2}{(a_3-a_2)^2 \a_2 \a_3 (\a_1+\a_2+\a_3)} \,, \ \ \ \ \ \ \ \ \ 
\t_2= \frac{1}{2}
\ln \frac{a_3^2 \a_1 \a_3(\a_1+\a_2+\a_3)}{ \a_2  (\a_3 a_3 - (\a_1+\a_3)a_2)^2}\,, 
\nonumber \\
&&\ \ \ \ \ \ \ \ \ \ \ \ \ \ \ \ \ \ \ \ \ \ \ \ \ \ \ 
\t_3= \frac{1}{2}
\ln \frac{a_2^2 \a_1 \a_2(\a_1+\a_2+\a_3)}{ \a_3  (\a_2 a_2 - (\a_1+\a_2)a_3)^2} \,.
\label{4.23}
\eea
Here  we defined the solution  using open string picture of Figure 2. 
To get the closed-string solution we are simply to double the $\s$-range
(the solution is  obviously periodic as it does not depend on $\s$).  

Finally, to get a candidate  solution of \rf{4.17},\rf{4.18} we  need 
to apply to  ~\eqref{4.23}  the  transformation~\eqref{4.16}
to map it to the complex $\xi$ plane with three marked points.
Note that as in the case of the 2-point function in section 3.2, 
all the information about the
points $\xi_i$  is hidden in this  Schwarz-Christoffel  map. 
To verify that the resulting $z(\xi), \ x(\xi)$ do  solve ~\eqref{4.17}, \eqref{4.18}
we may do this separately for the  three regions in \eqref{4.23}.
In region  I  we have  $\tau \in (-\infty, 0]$ and 
$\xi$ cannot reach the points $\xi_2, \xi_3$, i.e.  $\delta^2 (\xi-\xi_2)=\delta^2 (\xi-\xi_3)=0$. 
Then comparing the  r.h.s.  of eq.~\eqref{4.17} 
$
\frac{2 \pi \Delta_1}{b_1 \sqrt{\lambda}} \d^2 (\xi-\xi_1)\,
$
%
%
to its l.h.s. 
$ \frac{4 \k_1}{b_1}  \bar \pt \pt   \tau = \frac{2 \pi \k_1}{b_1}\d^2 (\xi-\xi_1)$
%
we conclude that 
$
\k_1= \frac{\Delta_1}{\sqrt{\lambda}}$.
%
The  regions II, III can be analysed in a similar way
implying that eq.~\eqref{4.17} is satisfied provided
\be
\kappa_2=\kappa_3 = \frac{\Delta_1}{\sqrt{\lambda}}=\k_1\,. 
\label{4.27} 
\ee
%
One can also  verify eq. \rf{4.18} as in the 2-point function case (see \rf{3.9},\rf{3.10}).

Finally, let us  compute the stationary-point value of  the $AdS_5$ part of the action
in \rf{4.4}. The string  part of the action   may be written as 
\bea
A_{0 AdS}= \frac{\sqrt{\lambda}}{\pi} \int d^2 \xi\ \frac{1}{z^2} (\partial z \bar \partial z+ 
\partial x \bar \partial x) 
= \frac{\sqrt{\lambda}}{\pi}\k_1^2  \int_I d^2 \xi \ \partial \tau \bar \partial \tau \ ,  
\label{4.28}
\eea
%
where $\tau$ is given by \rf{4.16}.
Integrating by parts and subtracting trivial  divergences we get
\bea
A_{0 AdS}= \frac{1}{\sqrt{\lambda}}\Big( 
{\D_1 \D_2} \ln |\xi_1-\xi_2|+ 
{\D_1 \D_3} \ln |\xi_1-\xi_3|-
{\D_2 \D_3} \ln |\xi_2-\xi_3|\Big)\,.
\label{4.29}
\eea
The term  in \rf{4.4} 
 involving  vertex operators is  straightforward to evaluate using the 
expressions for $\tau_1$, $\tau_2$, $\tau_3$ in~\eqref{4.23}: 
\bea
&&A'_{ AdS}= A_{AdS}- A_{0 AdS}=-\frac{2}{\sqrt{\lambda}}\Big( 
{\D_1 \D_2} \ln |\xi_1-\xi_2|+ 
{\D_1 \D_3} \ln |\xi_1-\xi_3|-
{\D_2 \D_3} \ln |\xi_2-\xi_3|\Big)
\nonumber \\
&&
\ \ \ +\ \D_1 \ln \frac{a_2 a_3}{a_3-a_2}
+\D_2 \ln \frac{a_2 (a_3-a_2)}{a_3}+  
\D_3 \ln \frac{a_3 (a_3-a_2)}{a_2}
\label{4.30}
\\
&&
\   -\  \frac{\D_1}{2} 
\ln  \frac{\a_2 \a_3 (\a_1 +\a_2 +\a_3)}{\a_1 (\a_2+\a_3)^2}
-\frac{\D_2}{2} 
\ln  \frac{\a_1 \a_3 (\a_1 +\a_2 +\a_3)}{\a_2 (\a_1+\a_3)^2}
 -\frac{\D_3}{2} 
\ln  \frac{\a_1 \a_2 (\a_1 +\a_2 +\a_3)}{\a_3 (\a_1+\a_2)^2}
\nonumber 
\eea
Summing up \rf{4.29} and \rf{4.30}  to get $A_{AdS}$ and adding also the $S^5$ part of the action in 
\eqref{4.15}  we obtain for the  leading  semiclassical  term 
in  the  3-point function \rf{4.2}
\bea &&  
G(a_1=0, a_2, a_3) = { 1 \ov \Omega_M} \int d^2 \xi_1 d^2 \xi_2 d^2 \xi_3 \ \  \GG (a_i; \xi_k)  \ ,
\la{gg}
\\ 
&& 
\GG \ \sim\ e^{-A_{AdS}-A_{S^5}} =\frac{C}{a_2^{\a_3} a_3^{\a_2} 
(a_3-a_2)^{\a_1} } e^{-\hat A(\xi_1, \xi_2, \xi_3)}\,, 
\label{4.31}
\eea
where $C$ is the same as in the supergravity expression in \rf{1.21} 
 (with $\a_i$ defined in \rf{1.19}; in the extremal case $C_{S^5}=1$)
\be 
C=C_{AdS} C_{S^5}=  \Big[ \frac{\a_1^{\a_1} \a_2^{\a_2}  \a_3^{\a_3} 
(\a_1+\a_2+\a_3)^{\a_1+\a_2+\a_3}}{(\a_1+\a_2)^{\a_1+\a_2} (\a_1+\a_3)^{\a_1+\a_3} 
(\a_2+\a_3)^{\a_2+\a_3}} \Big]^{1/2} \ . 
\label{4.32}
\ee
In the  extremal   case  under consideration $\a_1= \D_2+\D_3-\D_1=0$   so 
that finds that   $C=1$. 
The residual ``action''  $\hat A(\xi_1, \xi_2, \xi_3)$ is 
\bea
&&
\hat A(\xi_1, \xi_2, \xi_3) =
-\frac{1}{\sqrt{\lambda}}  (\D_1 \D_2 -J_1 J_2 )
\ln |\xi_1 -\xi_2| 
-\frac{1}{\sqrt{\lambda}} (\D_1 \D_3 -J_1 J_3 )
\ln |\xi_1 -\xi_3|
\nonumber \\
&&\ \ \ \ \ \ \ \ \ \ \ \ \ \ \ \ \ \ 
+\ \frac{1}{\sqrt{\lambda}}  (\D_2 \D_3 -J_2 J_3 )
\ln |\xi_2 -\xi_3|\,.
\label{4.33}
\eea
which vanishes due to the  marginality condition $\D_i=J_i$.

As in  the 2-point  function case in section 3.2, 
$\GG$ in \rf{4.31} contains also 
   an additional 
subleading contribution $ |\xi_1-\xi_2|^{-2}|\xi_1-\xi_3|^{-2} |\xi_2-\xi_3|^{-2}$  
 coming from  the 2-derivative factors in $\VV$ in the vertex operators 
\rf{41};   
 after the integration over $\xi_1,\xi_2,\xi_3$ cancels against the 
Mobius group volume  factor in \rf{gg} 
  as discussed in section 3.1  and Appendix A. 

%
%
The  final answer for the extremal ($\a_1=0$)  3-point function  \rf{4.2},\rf{4.31}
has  thus the expected ``factorized'' form 
(here we restore the  $a_1$ dependence)
\be
G (a_1, a_2, a_3) = 
\frac{1}{(a_1-a_2)^{\a_3} (a_1-a_3)^{\a_2} 
(a_3-a_2)^{\a_1} } = \frac{1}{(a_1-a_2)^{2\D_2} (a_1-a_3)^{2\D_3} }  \ . 
\label{4.35}
\ee
%



\section{Semiclassical computation of  non-extremal\\ three-point function}


Let us now consider  the case of generic $\D_i$.   
Here we shall start first with construction of semiclassical solution 
in the $AdS$ part. Our discussion in section 5.1 
will apply to  the case of generic
 {\it non-BPS}  string  states that 
carry  large charges in $S^5$ only so that the  relevant  part of the 
$AdS$ dependence of the vertex operators is the same as  in \rf{2.3},\rf{41}, i.e. $K^\D$. 
Then  the semiclassical trajectory   will be given again 
by 3 intersecting geodesics  but the Schwarz-Christoffel map will be more complicated
than \rf{4.16} as $\D_1$  is no longer   equal to $\D_2+\D_3$. 
In considering the $S^5$  contribution  in section 5.2 we shall specify to a   
non-extremal ($J_1 \not=J_2+J_3$)   case of 3  BPS operators. 
The  final semiclassical result for the 3-point correlator will match, of course, 
the supergravity   expression in \rf{1.21.00},\rf{1.32}.

\subsection{Solution in $AdS$}

The equations which we need to solve 
to find semiclassical trajectory in $AdS$ 
are still  the same as in ~\eqref{4.17},\eqref{4.18}.
One may expect that  the solution may still be given by 3 intersecting  geodesics in \eqref{4.22}
assuming the  Schwarz-Christoffel map  from a $(\tau,\sigma)$ domain 
to $\xi$ plane  and  the  corresponding regions I,II,III  are properly  defined. 
The expectation  that 
 the solution should  still be a function of one variable $\tau$
 is  supported by the following reasoning. 
 For semiclassical  string states that do not carry large charges in $AdS$
 the corresponding  $AdS_5$ solution should be  the same as for point-like BPS states 
 whose   correlation function is  reproduced   by the supergravity expression. 
The difference between  the BPS and non-BPS cases should  be 
visible only in the $S^5$ part of the semiclassical solution.

To construct  the relevant Schwarz-Christiffel map  let us  start with the conserved and traceless
(i.e. holomorphic)
  stress tensor of the $AdS$ part of the  classical string  sigma model in conformal gauge 
\be 
 T(\xi)\equiv T_{\xi\xi}= \frac{1}{z^2} [(\pt x)^2 + (\pt z)^2]\,.
\label{5.1}
\ee
If we assume that the required semiclassical  solution is given by~\eqref{4.22}
with  some choice of regions I,II,III  then computing $T$ in ~\eqref{5.1}  gives 
\be
T(\xi)=\kappa^2 ({\del \tau\ov \del \xi})^2\,,\ \ \ \ \ \ \ \ \ \ \k_1=\k_2=\k_3 \equiv \kappa \ ,
\label{5.6}
\ee
where  to make  $T(\xi)$  globally defined we have  to set 
$\k_i$ in~\eqref{4.22} to be equal.
Thus  to find  the map from the $\xi$-plane with 3 punctures to a $(\tau,\s)$ domain we need 
to know the exact form of $T(\xi)$.  

The key observation   is that  the structure of $T$
 can be fixed  uniquely \ci{j}  by  using (i) its expected 
 behavior near each marked point and  (ii) the conformal  transformation law
\be 
T (\xi; \xi_1, \xi_2, \xi_3) = \xi^4 T(\xi^{-1}; \xi_1^{-1}, \xi_2^{-1}, \xi_3^{-1})\,.
\label{5.2}
\ee
The behavior near each marked point is determined by the 2-point function solution~\eqref{3.6}
where $\tau$ is given by the conformal map~\eqref{3.5}.
Substituting this solution  into~\eqref{5.1} gives
\be
[T(\xi)]_{_{\rm 2-point}}= \kappa^2 (\pt \t)^2 =
\frac{\D^2}{4 \lambda} \frac{(\xi_1-\xi_2)^2}{(\xi-\xi_1)^2 (\xi-\xi_2)^2}\,, 
\label{5.3} 
\ee
where we used the conformal map~\eqref{3.5}. This means that near each marked point 
$\xi=\xi_i$  ($i=1,2,3$) the stress-energy tensor has to behave as
\be
T(\xi\to \xi_i) = \frac{d_i^2}{4} \frac{1}{(\xi-\xi_i)^2}\,, 
\qquad\qquad d_i \equiv \frac{\D_i}{\sqrt{\lambda}}\,.
\label{5.4}
\ee
Using \rf{5.2}  then allows one  to restore the exact form of $T$ 
%
\be
T(\xi)= \frac{d_1^2 (\xi_1-\xi_2)(\xi_1-\xi_3)}{4(\xi-\xi_1)^2 (\xi-\xi_2)(\xi-\xi_3)}
+
 \frac{d_2^2 (\xi_1-\xi_2)(\xi_3-\xi_2)}{4(\xi-\xi_1) (\xi-\xi_2)^2(\xi-\xi_3)}
+
\frac{d_3^2 (\xi_1-\xi_3)(\xi_2-\xi_3)}{4(\xi-\xi_1) (\xi-\xi_2)(\xi-\xi_3)^2}\,.
\label{5.5}
\ee
Comparing eqs.~\eqref{5.5} and~\eqref{5.6}
we  conclude that the required map is given by\foot{We implicitly assume that an arbitrary integration constant 
 can be absorbed into a shift of the origin of $\tau+ i \sigma$.}
%
\bea
\tau+ i \sigma =\frac{2}{\kappa} \int d \xi \sqrt{T(\xi)} \ , 
 \ \ \ \ \ \ \ {\rm i.e.} \ \  \ \ \ \ \ \ 
\tau= \frac{1}{\kappa}\int d \xi \sqrt{T(\xi)} + \frac{1}{\kappa}\int d \bar \xi \sqrt{\bar T(\bar \xi)} \ .
\label{5.7}
\eea
Eq.~\eqref{5.7} with $T$ given  by \rf{5.5}
 defines a new Schwarz-Christoffel map (with explicit form  given
 in Appendix B)
  that generalizes \rf{4.16}
 to the  generic  case of  $\D_1$ not necessarily equal
  to $\D_2+\D_3$.\foot{As already mentioned above, 
 we always choose $\D_1$ to  be the largest of the three  dimensions.} 
 Indeed,  in the extremal case
   $d_1=d_2+d_3$  the stress tensor \rf{5.5} simplifies to  
%
\be
T(\xi)= {1 \ov 4} \Big[ \frac{d_2  (\xi_1-\xi_2) } { (\xi-\xi_1)   (\xi-\xi_2) }
    + \frac{d_3  (\xi_1-\xi_3)}
{ (\xi-\xi_1)    (\xi-\xi_3)}\Big]^2
\label{5.8}
\ee
so that \rf{5.7}  implies that\foot{We ignore again an integration constant
that can be chosen to set, e.g.,   $\t_{int}=0$.}
\be
\tau+ i \sigma =\frac{1}{\kappa} \Big[ 
d_1  \ln (\xi-\xi_1) - d_2  \ln (\xi-\xi_2)- d_3 \ln (\xi-\xi_3) \Big] \ . 
\label{5.9}
\ee
This is equivalent to the map  \rf{4.6} used in the previous section 
if we set
 \be \kappa =d_1 \la{d11} 
\ee  
as we shall assume below.
In another special case considered  in \ci{ja}   when 
\be 
\xi_1 =\infty \ , \ \ \  \xi_2= 1 \ , \ \ \ \ \xi_3=-1 \ , \ \ \ \ \ \ \  \ \ \ \
d_2=d_3 \ , \la{s}
\ee
the stress tensor \rf{5.5} simplifies to 
\be 
T= { d_1^2 ( \xi^2 - q^2)  \ov 4(\xi^2 -1)^2} \ , \ \ \ \ \ \ \ \  \ \ \ \ \ 
 q^2 \equiv { d_1^2 - 4 d_2^2 \ov
d_1^2}  \ , \la{t}
\ee
and thus the map \rf{5.7} takes
 the form\foot{Note that in the extremal limit 
(when $q\to 0$)  this expression reduces to \rf{5.9} up to an irrelevant 
  divergent  constant $\sim \ln q$.}
\be 
\tau+ i \sigma = \ln (\xi + \sqrt{ \xi^2 - q^2})
   + {d_2 \ov d_1}  \Big( \ln {\xi-1 \ov \xi+1} 
   +  \ln {\xi + q^2  - \sqrt{1-q^2}  \sqrt{ \xi^2 - q^2} \ov 
  \xi - q^2  + \sqrt{1-q^2}  \sqrt{ \xi^2 - q^2}} \Big) \ . 
\label{jjj}
\ee

The discussion in~\eqref{5.5}, \eqref{5.7} is valid for arbitrary $\Delta_1$, $\Delta_2$, $\Delta_3$. 
However, the geometry of the complex domain in the $(\tau, \sigma)$ coordinates depends on the relation 
between the  $\Delta_i$'s. 
Let us now consider in more detail the  case when 
 $\Delta_1 > \D_2 +\D_3$ as then  it is easier to understand the 
structure of the  map \rf{5.7}.
 It is convenient again  to  view the closed-string  picture  with $\xi$  running over 
 a complex plane as a ``double'' of the 
open string   picture with $\xi$ belonging to the upper half plane
and $\xi_i$ lying on the real axis.
Then~\eqref{5.7} maps the upper 
half plane to the interior of a polygon on the complex $\tau+i \s$ 
plane  and which, in general,  is different  from the one in Figure 2. 
The critical points of the map~\eqref{5.7} are determined like in \rf{4.10} 
  from the equation 
$
{\partial  (\tau+ i \s) \ov \partial \xi} =0
$
(i.e. from zeroes of $T(\xi)$).
An important difference as compared to  the extremal case is that now this  equation
for $\xi=\xi_{int}$  is 
{\it quadratic} rather  than linear. The  resulting {\it two} solutions 
are given in~\eqref{A.3}, \eqref{A.4}. Note that for $\Delta_1 > \D_2 +\D_3$
the solutions in~\eqref{A.3}, \eqref{A.4} are {\it real} for $\xi_i$ lying along the real axis.
 Finding  $\tau$ and $\s$ on these two solutions\foot{In the special case of 
\rf{jjj}  we get $\xi_{int}= \pm q $ and thus 
$(\tau+i \sigma)_{int}=  \ln \xi_{int} 
+ {d_2\ov d_1}  \ln { (\xi_{int} -1) ( \xi_{int} +q^2) \ov (\xi_{int} +1) ( \xi_{int} -q^2) } 
$, 
i.e.  $ (\tau+i \sigma)_{int}^{(1)}= 
{1 \ov 2}  \ln { d_1^2 - 4 d_2^2 \ov d_1^2} + i \pi { d_2\ov d_1} \ , \ \ 
(\tau+i \sigma)_{int}^{(2)}= 
{1 \ov 2}  \ln { d_1^2 - 4 d_2^2 \ov d_1^2} + i \pi (1 -{ d_2\ov d_1}).$
} 
we get the same value for $\tau=\tau_{int}$ (which can be shifted 
to $\tau=0$), i.e. 
\be
\tau_{int}^{(1)}=\tau_{int}^{(2)}=0\,, 
\label{5.11}
\ee
while for  $\sigma=\s_{int}$ we get two different values 
\be
\s_{int}^{(1)} =\frac{\D_2}{\D_1} \pi\,, \qquad\qquad \s_{int}^{(2)} = (1 -\frac{\D_3}{\D_1})  \pi\,.
\label{5.12}
\ee
It is then  straightforward to draw 
the complex domain in $(\tau, \s)$ coordinates 
which is mapped to the upper half plane using~\eqref{5.7}
(see Figure 4). 
The left and right  ends of the three strips there are supposed to 
 run to infinity. 
The vertical size  of the ``removed'' region is given by 
$\s_{int}^{(2)} - \s_{int}^{(1)} =  \frac{\D_1-\D_2-\D_3 }{\D_1} \pi $, 
i.e. it vanishes in  the extremal case when  we get back to the diagram in Figure 2. 
\begin{figure}[ht]
\centering
\includegraphics[width=80mm]{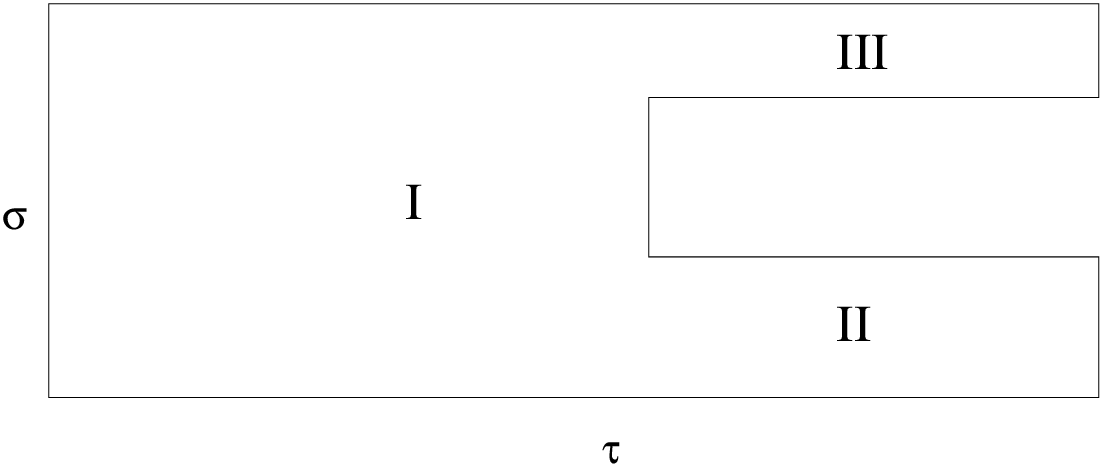}
\caption{\small The polygon on the complex plane $\tau + i \sigma$ 
whose interior is mapped to the upper half plane using the Schwarz-Christoffel map~\eqref{5.7}. 
The regions I,II,III correspond to  three interacting strings. 
}
\label{fig4}
\end{figure}
%
\noindent
The behavior near the interaction points  in 
 Figure 4 can be understood by an 
application of the 
Schwarz--Christoffel theorem (see, for example, ~\cite{Sid}) 
according to which  a general map from the upper half plane 
to the interior of a polygon is given by 
\be
\xi^{\prime} = \int d \xi\ (\xi-\eta_1)^{\d_1 -1} (\xi-\eta_2)^{\d_2 -1} \dots (\xi-\eta_n)^{\d_n -1} \,.
\label{zh1}
\ee
Here $\xi$ parametrizes the upper half plane, $\xi^{\prime}$ parametrizes
the interior of a polygon, $\eta_1, \dots \eta_n$ are the points along the real 
axis which are mapped to the vertices of the polygon and 
$\pi \d_1, \dots \pi \d_n$ are the angles at the corresponding vertices. 
Expressing $T$ in \rf{5.5}  in terms of the  critical points
 $\xi_{int}^{(1)}$ and $\xi_{int}^{(2)}$  of  the  map~\eqref{5.7}
(see ~\eqref{A.3}, \eqref{A.4})  we find 
\be
\sqrt{T}=\frac{(\xi-\xi_{int}^{(1)})^{1/2}\ (\xi-\xi_{int}^{(2)})^{1/2}}{(\xi-\xi_1)
(\xi-\xi_2)(\xi-\xi_3)}\,.
\label{zh2}
\ee
Comparing with~\rf{5.7} with \eqref{zh1} we  conclude that
in our case $\eta_{i}=\xi_i, \ i=1,2,3$ with $\delta_i=0$ 
and $\eta_{4,5}= \xi_{int}^{(1,2)}$ with $\delta_i={3 \ov 2}$, 
i.e. 
 the angles at the interaction points
are $3 \pi\ov 2$ as, indeed, shown on  Figure 4.
 Note that in the extremal case the points
$\xi_{int}^{(1)}$ and $\xi_{int}^{(2)}$ coincide  and near   this point 
$\sqrt{T}$ behaves as $\xi-\xi_{int}$,  in the corresponding angle  is 
$ 2 \pi$ in agreement with Figure 2.

Comparing Figure 2 and Figure 4   one may be formally interpret the latter as 
 corresponding to a ``generalized''  light-cone interacting string diagram 
 where  $p^+$ momentum, i.e.  length of the string, is not conserved:
 the ``removed'' region in Figure 4  may stand for an external state  (carrying away 
 the deficit of momentum or $\D_1 - \D_2- \D_3$ in the present context).

Clearly, the  Figure 4 applies  to the case  when 
$\Delta_1 > \D_2 +\D_3$. 
In the opposite case ~\eqref{5.7}  is not defined as a map from the upper half plane. 
The reason is that for $\xi_i$ lying along the real axis the critical points
are always complex with non-zero imaginary part 
(see ~\eqref{A.3}, \eqref{A.4}), i.e. 
in this case we cannot view the resulting closed string worldsheet 
as two copies of a polygon with proper identifications along $\sigma$. 
Then  we have to  interpret ~\eqref{5.7} as a map from the full complex plane and 
the  resulting  $(\tau, \sigma)$ domain and the individual regions I, II, III are harder to 
visualise.\footnote{Note
 that in the case of the three BPS operators one always has 
$\Delta_1\leq \Delta_2+\Delta_3$, $\Delta_2\leq  \Delta_1+\Delta_2$, $\Delta_3\leq  \Delta_1+\Delta_2$ (this is obvious at weak coupling 
and holds in general due to non-renormalization). 
Thus, Figure 4 does not apply to 
the (non-extremal) three-point function of the three  BPS operators
(we  thank G. Georgiou for pointing this out to us). 
Nevertheless, since the geometry of the domain in the $(\tau, \sigma)$ coordinates
is simpler for $\Delta_1 > \Delta_2 +\Delta_3$ it is  convenient to formally perform the analysis in this case, treating the opposite case by 
analytic continuation. 
The general map~\eqref{5.7} and the final results are indeed  valid for
 arbitrary $\Delta_i$'s.}

The proposed solution  to eqs.~\eqref{4.17}, \eqref{4.18} is thus 
given by the expressions in \eqref{4.22}  where 
the regions I, II, III  should  now  be defined (for $\Delta_1 > \Delta_2 +\Delta_3$) as in Figure 4.
For example, let us consider eq.~\eqref{4.17}. 
In each of the three regions  only one marked point $\xi_i$  is contributing. 
Just like in the previous section, 
near each marked point  the l.h.s. and the r.h.s. of \eqref{4.17} are equal to 
each other 
\be
\frac{4 \kappa}{|a_i -b_i|} \pt \bar \pt  \tau=
\pm \frac{2 \pi d_i}{|a_i -b_i|} \d^2 (\xi-\xi_i)\,, 
\label{5.13}
\ee
where the choice of the sign depends whether the point $\xi_i$ is mapped to 
$\t=\infty$ or $\t=-\infty$.  
For concreteness, we choose the convention that
$\xi_1$ is mapped to 
$-\infty$ and $\xi_2$, $\xi_3$ are mapped to $+\infty$.
Eq.~\eqref{5.13} follows from the fact that  according to \rf{5.6},\rf{5.7}  near each puncture 
  $\kappa \pt \tau=\sqrt{T}$ 
 has a simple pole $\sim (\xi-\xi_i)^{-1}$ 
with residue $\pm d_i$.

 The calculation of the  corresponding semiclassical  value of the $AdS$ part of the action 
 in \rf{4.4}  is the same as in the previous section (see \rf{4.29},\rf{4.30})
 and we will simply state the result (restoring  the dependence on $a_1$)
\be
e^{-A_{AdS}} =\frac{C_{0 AdS}}{(a_2-a_1)^{\a_3} (a_3-a_1)^{\a_2}(a_3-a_2)^{\a_1}}
\ e^{-\hat A_{AdS} (\xi_1, \xi_2, \xi_3)}\,,
\label{5.14}\\
\ee
where,
$C_{0 AdS}$ is the same as $C$ in \eqref{4.32}, i.e. 
\be 
C_{0 AdS}= \Big[ \frac{\a_1^{\a_1} \a_2^{\a_2}  \a_3^{\a_3} 
(\a_1+\a_2+\a_3)^{\a_1+\a_2+\a_3}}{(\a_1+\a_2)^{\a_1+\a_2} (\a_1
+\a_3)^{\a_1+\a_3} (\a_2+\a_3)^{\a_2+\a_3}} \Big]^{1/2}
\ , \label{5.15}
\ee
and $\hat A_{AdS}$
\be
\hat A_{AdS}(\xi_1, \xi_2, \xi_3)= \frac{\k}{2} \int d^2 \xi\  \Big[\D_1 \d^2 (\xi-\xi_1) - 
\D_2  \d^2 (\xi-\xi_2) -\D_3  \d^2 (\xi-\xi_3) \Big]  \t(\xi,\bar \xi) \,.
\label{5.16}
\ee
When one substitutes here $\tau$ computed using \rf{5.7},\rf{A.1}  one finds 3 types of terms: 
(i)  divergent ``self-contraction'' terms that should be subtracted; (ii) 
$\ln|\xi_i-\xi_j|$ terms  that   will cancel against similar $S^5$
terms after use of  marginality condition as in   \rf{4.33};
(iii)   $\xi_i$-independent 
terms $\sim \D_i \ln \D_j$ which contribute  an extra   factor  $C'_{AdS}$
  to the structure constant  in the 3-point function, i.e.  
 $C_{AdS}=C_{0 AdS}\ C'_{AdS}$. 
To  compute $C'_{AdS}$ using \rf{A.1},\rf{A.2} 
 one is to take into account  the choice of $\tau_{int}=0$
 which means  that $\tau$ is to be shifted  by the following constant 
%
\be
\hat{\tau}=\frac{d_1-d_2-d_3}{2d_1}\Big(
 \ln \big[d_1^4+ (d_2-d_3)^2 -2 d_1^2 (d_2^2 +d_3^2)\big] + \ln \big[
|\xi_1-\xi_2||\xi_1-\xi_3||\xi_2-\xi_3|\big]\Big) \,.
\label{5.16.1}
\ee
Then the additional contribution  coming from~\eqref{5.16}  is found to be 
\bea
&&
\ln C'_{AdS}=\frac{1}{2}\sqrt{\lambda}\Big[ - d_1^2 \ln (4 d_1^2)
+ d_2^2 \ln (4 d_2^2) + d_3^2 \ln (4 d_3^2) 
\nonumber \\
&&
\ \ \ +\ 2d_1 d_2 \ln\big[(d_1+d_2-d_3)(d_1+d_2+d_3)\big]
+2d_1 d_3 \ln\big[(d_1+d_3-d_2)(d_1+d_2+d_3)\big]
\no \\
&&
\ \ \ -\  2 d_2 d_3 \ln\big[(d_1+d_2-d_3)(d_1+d_3-d_2)\big]
 \label{5.16.2}\\ &&
\ \  \ + \  {1 \ov 2}  (d_1 - d_2 -d_3)^2 \ln  \big(d_1^4+ d_2^4  +d_3^4  -2 d_1^2 d_2^2 -2 d_3^2 d_2^2 
-2 d_3^2 d_1^2 \big) \Big] \no 
\eea
As we shall see in the next subsection,  in the BPS case
this additional   contribution cancels 
against a similar  contribution  coming from $S^5$
(like in the supergravity  approach in section 2.3 and in the
extremal case  in \rf{4.33}). The reason   for this cancellation 
can be traced  to the marginality  condition that ``links'' the $AdS_5$ and $S^5$ 
contribution.\foot{Such cancellation may happen also in  more general   
context, as it is linked with cancellation of $\ln|\xi_i-\xi_j|$ terms 
that should have only  ``subleading''  (i.e. not proportional to $\sqrt \lambda$) 
coefficients in order to ensure consistency with 2d conformal invariance
(and, in particular, cancellation of Mobius volume factor).}

\subsection{Solution in $S^5$ for non-extremal BPS  correlator }


The $S^5$  contribution depends on a particular choice of the vertex 
operators. In this section we will consider the case of  all three operators being BPS
and choose them so that they represent a  non-extremal  correlator. 

Like in sections 2.3, 2.4   we may  first consider  special case and then generalize. 
 Namely, let us start with  the 
 same  $S^5$   wave functions as in  \rf{1.5.5},\rf{1.22},\rf{1.23}
\be
v_1(\xi_1)= (X_1+ i X_2)^{J_1}\,,  \ \ \ \ \ \ 
v_2(\xi_2)=  (X_1- i X_2)^{J_2}\,, \ \ \ \ \ \ \ 
v_3(\xi_3)=  (X_1+ i X_3)^{J_3}\,.
\label{5.18}
\ee
Introducing the angles $\varphi$ and $\psi$ as in  \rf{1.24}
 we arrive at the following  $S^5$ part of the effective action including the relevant 
 (large-charge)  parts of the vertex operators
\bea
A_{S^5} &=& \frac{\sqrt{\lambda}}{\pi} \int d^2 \xi\ \big(\pt \psi \bar \pt \psi + \cos^2 \psi \pt \varphi
\bar \pt \varphi\big)
\nonumber \\
& &  \ \ \ - \  J_1 \int d^2 \xi\ \d^2 (\xi-\xi_1) \ln (\cos \psi e^{i \varphi}) -J_2 
\int d^2 \xi\ \d^2 (\xi-\xi_2) \ln (\cos \psi e^{-i \varphi})
\nonumber \\
& & \ \ \ - \ J_3
\int d^2 \xi\ \d^2 (\xi-\xi_3) \ln (\cos \psi \cos \varphi + i \sin \psi)\,,
\label{5.19}
\eea
where the  first term ($S^5$ part of string action)
we ignored  all the fields that vanish on the semiclassical trajectory.
 The  analysis in section 2 suggests that it is useful  to perform the 
analytic continuation \rf{1.25}, i.e. 
\be
i X_2 \to \tX_2 \,, \quad i X_3 \to \tX_3 \,, \qquad {\rm i.e.} \qquad 
i \varphi \to \tvarphi \,,\quad i \psi \to \tpsi \,.
\label{5.19.1}
\ee
Then from~\eqref{5.19} we obtain the following equations of motion 
\bea
&&
 2\pt \bar \pt \psi -\sinh 2 \psi\ \pt \varphi 
\bar \pt \varphi
\nonumber \\
&&\ \ =  \frac{\pi}{\sqrt{\lambda} }\Big[
J_1 \tanh \tpsi\ \d^2 (\xi-\xi_1) +J_2 \tanh \tpsi\ \d^2 (\xi-\xi_2) 
+J_3 \frac{\tanh \tpsi \cosh \tvarphi +1}{\cosh \tvarphi +\tanh \tpsi}\ \d^2 (\xi-\xi_3)\Big]
\,, 
\nonumber \\
&&
\pt (\cos^2 \tpsi\ \bar \pt \tvarphi) 
+\bar \pt (\cos^2 \tpsi\  \pt \tvarphi)
\nonumber\\
&&\ \ = \frac{\pi}{\sqrt{\lambda} }\Big[
J_1\ \d^2 (\xi-\xi_1) - J_2\ \d^2 (\xi-\xi_2) +J_3 \frac{\sinh \tvarphi}{\cosh \tvarphi +\tanh
\tpsi}\ 
\d^2 (\xi-\xi_3)\Big]\,.
\label{5.20}
\eea
As in the discussion of the  $AdS$  case 
we have to impose the condition that there is no additional singularity 
at $\xi=\infty$. This gives us eqs.~\eqref{1.27} whose solution is given in~\eqref{1.28}.
The problem  then 
is how to construct the local solutions in the regions I,II,III 
and glue them at the point~\eqref{1.28} at $\tau=0$.
 Since we are considering BPS operators the local solutions
must be again geodesics. 
Naively, one might  think 
that the relevant solutions should  be simply (as in  \rf{4.8}) given by  
$\tvarphi =\kappa \tau\,, \  \tpsi =0$  and $\tpsi=\kappa \tau\,, \  \tvarphi =0$
%
%
but these cannot be glued at~\eqref{1.28}. The right choice of  
(complexified)
 geodesics in regions I,II,III
is more complicated. 
Fortunately, as we discussed at the end of subsection 2.3, 
 we can reduce the  problem  of finding them 
 to  an equivalent one in $AdS_2$ 
and thus simply borrow the results from the previous subsection!

Explicitly,     the analytic continuation~\eqref{5.19.1}  maps 
the sphere $X_1^2 + X_2^2 + X_3^2=1$ 
 into the euclidean $AdS_2$ space $X_1^2 - \tX_2^2 -\tX_3^2=1$.
   Introducing there  
the Poincare coordinates  $(r,y)$ ~\eqref{1.32.1} so
that the original $S^2$ angles  $( \varphi,\psi)$ are given by 
\be \la{an}
e^{ 2 i \varphi} = { r^2 + (y+1)^2 \ov r^2 + (y-1)^2} \ , \ \ \ \ \  \ \ \ \ \  \ \ \ \ \ 
\sinh (i \psi) =  { r^2 + y^2 -1 \ov 2r}  \ , 
\ee
we  get for  the vertex operator factors in \rf{5.18} 
\bea
&&v_1 = \frac{1}{2^{J_1}} \Big( \frac{r}{r^2 +(y+1)^2}\Big)^{-J_1}, \ \ \ \ \ \ \ \ \ \ \ \ 
v_2 = \frac{1}{2^{J_2}} \Big( \frac{r}{r^2 +(y-1)^2}\Big)^{-J_2},\no \\   &&
\ \ \ \ \ \ \ \ \ \ \ \ \ \ \ \ \ \ \ \ \ \ \ \ \ 
v_3  =\Big( \frac{r}{r^2 +y^2}\Big)^{-J_3}\,.
\label{5.22}
\eea
These may be formally 
 interpreted as vertex operators in $AdS_2$ inserted at  the boundary points 
$a_1=-1$, $a_2=1$, $a_3=0$ and carrying effective dimensions $-J_1$, $-J_2$, $-J_3$. 
The corresponding  semiclassical solution  can thus be found from \rf{4.20},\rf{4.21},\rf{4.22}
where  one is  to replace $(z,x) \to (r,y)$ and also 
to interchange the points $a_1$ and $a_3$
 (as we assumed in \rf{4.20}  that $a_1=0$). Its explicit $S^2$ form can then be written 
  using \rf{an}, i.e. this solution is  complex in terms of the original  coordinates.
  The fact that the $S^5$ intersection point is complex  was already found in 
  \rf{1.28}.\foot{This solution  is thus different from the $S^5$ part of the 
  3-geodesic solution discussed in \ci{km}.}

The action 
 on this  solution
 was  already found in~\eqref{5.14}--\eqref{5.16}  so we should just 
substitute the above data
(we should also remember to include the factor $2^{-J_1-J_2}$ 
coming from eqs.~\eqref{5.22}). 
As a result,  we obtain 
\bea
&&e^{-A_{S^5}} \ =\  C_{S^5}\  e^{-\hat A_{S^5} (\xi_1, \xi_2, \xi_3)}\,, 
\label{5.23} \\
&&
C_{S^5}= \frac{1}{2^{J_3}} \ 
\Big[ \frac{ (\b_1+\b_2)^{\b_1+\b_2} (\b_1+\b_3)^{\b_1+\b_3} (\b_2+\b_3)^{\b_2+\b_3}}
{\b_1^{\b_1}\b_2^{\b_2} \b_3^{\b_3} (\b_1+\b_2+\b_3)^{\b_1+\b_2+\b_3}}\Big]^{1/2} \ , 
\label{5.24}\\
&& 
\hat A_{S^5} (\xi_1, \xi_2, \xi_3) = 
\frac{\k }{2} \int d^2 \xi \Big[ - J_1  \d^2 (\xi-\xi_1) + 
  J_2 \ \d^2 (\xi-\xi_2)  +  J_3 \ \d^2 (\xi-\xi_3)\Big]  \ \t(\xi, \bar \xi)
\label{5.25}
\eea
where $\b_i$ were defined in \rf{1.30}.

Combining this  with the $AdS$  contribution in ~\eqref{5.14}--\eqref{5.16}  and using 
marginality condition
$\D_i=J_i$ 
we find that  
$\hat A_{AdS} $ cancels against $\hat A_{S^5}$. 
This implies, in particular, that  that $C'_{AdS}$ in~\eqref{5.16.2} indeed cancels out.\foot{Let us 
stress  that  
 this cancellation is not due to the  relation between $S^2$ and $AdS_2$
which appeared  because of the analytic continuation~\eqref{5.19.1}
but is  due to the simple marginality condition for the BPS operators.}
Since $\a_i=\b_i$  we find also that  $C_{AdS}$ cancels against the square root  factor 
in~\eqref{5.24}, i.e. we are  left with the same 3-point coefficient 
\be
C=\ C_{AdS}\ C_{S^5} \ =\frac{1}{2^{J_3}}
\label{5.26}
\ee
as found    in  the supergravity and  free gauge theory  computations in section 2. 

The discussion  of more general  case of non-extremal  correlators 
considered in section 2.4  is of course straightforward 
using again the analytic continuation to $AdS_5$.\foot{As was used 
above,   one is 
take into account that  under the analytic continuation from $S^5$ to $AdS_5$ 
one is to invert the sign  of the string action  so that the semiclassical solution 
remains the same with $\a_i \to \b_i$.}
 The resulting string theory expression is again the same as in 
 \rf{011}.


\section{
An example of semiclassical three-point function\\
  of non-BPS  operators}


In this section we will study an example of 3-point function 
of non-BPS states  that correspond to 
``small''  circular strings in $S^3$.\foot{Attempts to discuss more apparently subtle 
examples  with ``large'' circular  strings 
wrapping   big circle of $S^3$ were made  in \ci{km,ry}.
} 
If we parametrize  the 5-sphere as in ~\eqref{1.23.1}, i.e. 
%
\bea
&&
X_1+ i X_2 =\cos \theta \cos \psi e^{i \varphi_1}\,, 
\ \ \ \ 
X_3+ i X_4 =\cos \theta \sin \psi e^{i \varphi_2}\,, 
\ \ \ \ 
 X_5 + i X_6 =\sin \theta e^{i \varphi_3}\,,
\label{6.1}
\eea
then  the  classical solution representing a ``small''  circular 
string  rotating on  $S^3$  of radius $0 < a < 1 $ inside $S^5$  with two equal angular momenta 
has the following simple ``chiral'' form ($AdS$ time is $t=\k \tau$)  ~\cite{Multi}
\bea 
&&X_1+ i X_2=a\ e^{i (\tau + \s)} \ , \ \ \ \ \  X_3+ i X_4= a\ e^{i (\tau - \s)}\ , \ \ \ \ 
X_5 + i X_6= \sqrt{1 - 2 a^2} \ , \la{61} \\
&&   J_{12}=  J_{34} \equiv J = \sql a^2  \ , \ \ \ \ \ \ \ \ \ \ 
E= \sql  \k = 2 \sqrt{\sql J}  \ . \la{16}
\eea
The $AdS$ energy $E$  of this solution has exactly the same form 
as in flat space (with $\sql \to {1 \ov \a'}$)  where the string solution 
 described  by 4 cartesian coordinates is  given by\foot{This  configuration belongs to $S^3 \subset R^4$ and  thus  can be   directly embedded
 into $S^5$.} 
 $x_1+i x_2 =a e^{i (\tau + \s)}, \ \   x_3 + i x_4= a e^{i (\tau - \s)}$.
 
Since $a$ can be taken to be small,  
it is natural to  expect that the $S^5$ part of the 
vertex operator representing such state  should have  similar structure to its flat space
counterpart in $R_t \times R^4$ (in ``momentum'' representation) 
\be 
\int d^2 \xi \   e^{-i E t}  \  \big[\pt (x_1 + i x_2)\big] ^{J} \big[\bar \pt (x_3 + i x_4)
 \big]^{J}  \ ,  
\la{7}
\ee
i.e. 
 (cf. \rf{3.1})
\be 
\V(\vec a) = \int d^2 \xi \  \Big[ { z \ov z^2 + ( \vec x - \vec a)^2}\Big]^\D \   v(\xi)  \ , \ \ \
\ \ \  \ v(\xi)= \big[\pt (X_1 + i X_2)\big]^{J}\big[\bar \pt (X_3 + i X_4)\big]^{J}   \ . 
\la{67}
\ee
The semiclassical  approximation 
to the  2-point function of such  operators 
 is governed~\cite{E1} by the  geodesic in $AdS$ \rf{3.6} combined with 
 the euclidean continuation ($\tau \to -i \tau$)  of the  classical solution ~\rf{61}, i.e. 
\bea
&&
i \varphi_1 = \tau + i \s\,, \qquad 
i \varphi_2 = \tau - i \s\,,
\qquad\quad
\cos \theta=\sqrt 2 a\,, \qquad \psi=\frac{\pi}{4}\,, \qquad \varphi_3=0\,,
\label{6.4}
\eea
with 
\be \D=E = 2 \sqrt{\sql J} \la{m}  \ee
  being the marginality condition. 
This solution should be mapped to the complex $\xi$-plane with two marked points by the same 
 map as in \rf{3.5}, i.e. 
\be
\tau+ i \s =\ln (\xi-\xi_1)- \ln (\xi-\xi_2)\,.
\label{6.4.1}
\ee
Let  us now consider computing  a correlation function 
of the 3 operators like \rf{67}  in semiclassical  approximation 
assuming  $J_i \sim \sql \gg 1$ and  $\D_i = 2 \sqrt{\sql J_i} \sim \sql \gg 1$
  choosing  their $S^5$ parts  in  the following particular  form:
\bea &&
v_1(\xi_1) = \big[\pt (X_1 + i X_2)\big]^{J_1}\big[\bar \pt (X_3 + i X_4)\big]^{J_1} \ , \quad
v_2(\xi_2)= \big[\pt (X_1 - i X_2)\big]^{J_2}\big[\bar \pt (X_3 - i X_4)\big]^{J_2} \ , \no\\
&&\ \ \ \ \ \ \ \ \ \  \ \ \ \ \ \ \ \ \ \ v_3(\xi_3)= \big[\pt (X_1 - i X_2)\big]^{J_3}\big[
\bar \pt (X_3 - i X_4)\big]^{J_3} \ ,
\label{6.2}
\eea
Note that  with this choice all the three operators  correspond to strings 
spinning in the same $S^3$. 
In this case, as in flat space, 
  the integrals over the zero modes  of $\varphi_1 $ and $\varphi_2 $
appear to impose  angular momentum conservation constraint
%
\be
J_1= J_2+J_3\,. \label{6.3}
\ee
Then the corresponding correlator in flat space  will vanish  if restricted to 
$R_t \times R^4$ as \rf{6.3} with the mass shell condition \rf{m} 
will be inconsistent with the energy  conservation $E_1 = E_2 +E_3$.
To get a non-zero correlator we will  need enlarge phase space
introducing non-zero momentum components in other directions, so that  the flat-space
marginality conditions  become $E^2_i - \vec p^2_i = 4\a'^{-1}  J_i$.

Let us see what happens in the \adss case  were 
 there is no  a priori   conservation condition   for $\Delta_i$.
  The $AdS_5$ part of the semiclassical solution should be exactly as in  non-extremal case 
  discussed in section 5.1.
  As for the $S^5$ part of the solution, we will argue that it  
  given by~\eqref{6.4}  with 
  \be
\tau+i \s = \ln (\xi-\xi_1) - \frac{J_2}{J_1} \ln (\xi-\xi_2) - \frac{J_3}{J_1} \ln (\xi-\xi_3)
\label{6.9} \ . 
\ee
%
%
The form of this  map is suggested to be  the same as in 
 the extremal BPS case \rf{4.9} since $J_i$ are conserved
 and since the angles are linear in $\tau$ and $\sigma$ as in the flat space case.

The stationary-point 
equations of motion for $\varphi_3$, $\theta$, $\psi$  happen to be 
non-singular and are solved  by the same relations 
 $\cos \theta=\sqrt 2 a$, $\psi={\pi\ov 4}$, $\varphi_3=0$  as in~\eqref{6.4}  together 
with  the conditions that $\varphi_1$ is holomorphic and  $\varphi_2$ is antiholomorphic.
The equation for $\varphi_1$ reads
\bea
&&
\frac{\sqrt{\lambda} a^2 }{ \pi} (\pt \bar \pt + \bar \pt \pt) i\varphi_1=
J_1 \d^2 (\xi-\xi_1) -J_2 \d^2 (\xi-\xi_2)-J_3 \d^2 (\xi-\xi_3)
\nonumber\\
&&
- \pt \Big( ( \pt \ln i \varphi_1)^{-1}\big[
 J_1  \d^2 (\xi-\xi_1)-J_2 \d^2 (\xi-\xi_2)-J_3 \d^2 (\xi-\xi_3)\big] \Big)
\label{6.10}
\eea
and  the equation for $\varphi_2$ is obtained from~\eqref{6.10} by 
replacing $\varphi_1 \to \varphi_2$, $\pt \to \bar \pt$. 
Since on the solution~\eqref{6.4}  with  \eqref{6.9} 
one has 
$
(\d^2 (\xi-\xi_i))^{-1} {\pt \ln i \varphi_{1, 2}}=0, 
$
we find that eq.~\eqref{6.10} is indeed solved by~\eqref{6.4}, \eqref{6.10}
provided 
\be 
J_1= \sqrt{\lambda}\  a^2 \,.
\label{6.11}
\ee
The $S^5$ part of the string  stress tensor on this solution is found to be 
\bea
&&
T_{S^5} (\xi)= \cos^2 \theta \cos^2 \psi\ (\pt \varphi_1)^2 
=-\frac{1}{\sqrt{\lambda} J_1 }\Big[  \frac{J_2(\xi_1-\xi_2)}{(\xi-\xi_1)(\xi-\xi_2)}
  +  \frac{J_3(\xi_1-\xi_3)}
{(\xi-\xi_1)(\xi-\xi_3)}\Big]^2\,.
\label{6.12}
\eea
Conformal gauge condition   requires that the full \adss stress-energy tensor should  vanish. 
This means that~\eqref{6.12} has to cancel the $AdS_5$ contribution ~\eqref{5.5}
with $d_1^2 =d_2^2 +d_3^2$\  ($d_i\equiv { \D_i\ov \sql}$):
this relation   follows from the angular momentum 
conservation \rf{6.3} and the marginality condition  \rf{m}. 
  However, it is easy to see that 
this cancellation (cf. \rf{5.8})  and thus the agreement  between the 
$AdS_5$ map \rf{5.9} and the $S^5$ map \rf{6.9} 
is impossible:  it requires $d_1=d_2+d_3$
 in addition  to   $d_1^2 =d_2^2 +d_3^2$  implying $d_i=0$. 
This suggest that in this case  semiclassical  solution does not exist 
which we interpret as an indication that this  correlator should 
vanish  as in flat space. 

The clash between the angular momentum  conservation and  the nonlinear (non-BPS)
marginality condition can be avoided by considering  analogs of non extremal BPS correlators
discussed in the previous sections. There   the three operators  carry  charges
from different planes so that the  charge conservation applies only ``pair-wise''.
Semiclassical computation of  such correlators remains an interesting open problem.

\section{Concluding remarks
}


In this section we would make some comments on 
comparison of  our approach with  that of ref. 
~\cite{ja}.
The authors of ~\cite{ja} suggested a construction of the $AdS$  part of 
the semiclassical solution  corresponding to a  correlator of 3 
operators that carry large charges in $S^5$ only by using the Pohlmeyer reduction 
(see, e.g., ~\cite{Pohl})
to find the relevant $AdS_2$ solution.\foot{The boundary 
points $\vec a_i$  for the  3 operators  were assumed to lie on a line.}
They defined the reduced theory variable $\tilde{\g}$ by 
\be 
\frac{\pt z \bar \pt z + \pt x \bar \pt x }{z^2}= \sqrt{T \bar T} \cosh \tilde{\g}\,,
\label{7.1}
\ee
where $T$ is  the stress tensor  ~\eqref{5.5}
corresponding to  the case of the three  generic dimensions $\D_i$\footnote{As was mentioned earlier, 
in ~\cite{ja} the 
insertion points  and dimensions were chosen as in \rf{s}.
}
so that it satisfies a generalized sinh--Gordon equation
\be 
\pt \bar \pt \td{\g}= \sqrt{T \bar T} \sinh \tilde{\g}\,.
\label{7.2}
\ee
Given a solution  for $\td \g$,    to find the original Poincare coordinates 
$z,x$ one is to solve an additional linear problem  (see \cite{ja} for details).

In this framework, the solution which we suggested 
 in section 5.1 (that should apply to generic non-BPS operators with charges only in $S^5$)
is simply $\tilde{\gamma}=0$. 
In~\cite{ja} this  case  was 
excluded   as corresponding  to the geodesic related  to  the 2-point function
and it was   assumed that the 3-point correlator should be described 
 by a non-trivial  solution $\td \g \not=0$ of   \rf{7.2}.
 However,   $\tilde{\gamma}=0$
does not necessarily correspond just to the 2-point function since 
there is an additional data  associated to the 3-point function case.

Indeed, the  3-point function problem   is  defined  on  a plane with 
three punctures  rather than two. 
Using the Schwarz--Christoffel transformation defined by the stress tensor 
 we can map the plane with three marked points to a complex domain
in $(\tau, \s)$ plane. Part of  non-triviality of the 
solution is thus  hidden in the Schwarz--Christoffel map, i.e.  
in  details of the $(\tau, \s)$ domain. 
While the solution suggested in section 5.1 (which generalizes 
the 3-geodesic configuration of \ci{km} in the  BPS case) 
in each of the  three $(\tau, \s)$  regions  corresponds simply to 
 the $\tilde{\gamma}=0$ one as in the 2-point function  case,  
the gluing condition, i.e.    the precise definition
of the three regions depends on the Schwarz--Christoffel map and, 
hence, on the stress tensor. 
%


We believe  that for  given generic values of dimensions $\D_i$ 
the $AdS$ part of the semiclassical solution controlling the 3-point function 
 should be {\it the same}
in the case of  non-BPS  operators   
 as in the (non-extremal) case  of BPS  operators:  as the corresponding vertex operators 
 are assumed to carry 
  only  $S^5$ charges, 
 the distinction  between the two cases should be visible only in the $S^5$
 part of the semiclassical solution. At the same time, as we 
 demonstrated in this paper, the expected value of the
  BPS correlator is correctly  reproduced  by the ``point-like'' 
  3-geodesic solution \rf{4.22}.
 
 Ref. \cite{ja} claimed that  the relevant  $AdS$ solution 
 should  be described by a non-trivial $\td \g \not=0$ 
 and that  the   case  of the BPS correlator should be recovered only 
 in the case when  $d_i= {\D_i \ov \sql}$ are  small.
  This   formally follows from  \rf{7.2}  since  
  in view of \rf{5.5}  the coefficient $\sqrt{T \bar T}$  in \rf{7.2}
  is small for small  $d_i$  and thus the solution  of \rf{7.2}
  should be well approximated  by $\td \g =0$ one. 
  However,  the BPS states can, of course,  carry any large charges and 
  thus have $d_i \gg 1$   so we believe that the relevant  solution 
  of \rf{7.2} should   be just  $\td \g =0$ for any values of $d_i$.


There are,  obviously,   many  open problems. It remains 
to find a non-trivial example of non-BPS correlator with 
$S^5$ charges, i.e. to  construct the $S^5$ part 
of the corresponding solution. 
One should also address  the same question 
for  correlators with non-trivial  charges in $AdS_5$, 
generalizing the  approach  in \ci{ja} (for very recent 
work in this  direction see \ci{ka}).



\section*{Acknowledgments}
We  thank G. Georgiou, R. Janik, T. McLoughlin, T. Klose,  R. Roiban 
 and L. Uruchurtu for useful  discussions. 
The work of E.I.B. is supported by an STFC  grant.
 A.A.T. acknowledges the support of the RMHE grant 2011-1.5-508-004-016.


\appendix
\section{World-sheet and target-space  conformal 
 symmetry\\  factors in  string  correlation functions in $AdS$} 

Here we shall  explain in detail  the  remarks made in section 3.1   about the
symmetry group   factors  in the  2-point  and 3-point correlation functions 
in string theory in $AdS$ space. 
For concreteness, we will present  
the discussion   in the framework of semiclassical expansion  used in this  paper.

Let us start with   evaluating  the factor  $\oc$ in \rf{2.8} which is the volume 
of the subgroup of the Mobius group\foot{The corresponding volume can 
be written  as $\int d^2 a\ d^2 b\ d^2 c\ d^2 d {\ } \delta^2( ad-bc-1)$.}
\be
\xi^{\prime}= \frac{a \xi + b}{c \xi +d}\,, \qquad a, b, c, d \in {\mathbb C}\,, \qquad ad - bc =1\,, 
\label{2.10}
\ee
 preserving two points on a complex plane. 
We shall   choose these points 
to be  $0$ and  $\infty$, so that the  transformations preserving them  will have 
\be
b=0\ , \ \ \ \ \ \  \quad c=0\,,  \ \ \ \ \ \ \    \quad a=d^{-1} = r e^{i \theta}\,.
\label{2.11}
\ee
They thus consist of dilatations with parameter $r=|a|$ and  $U(1)$ rotations 
with parameter $\theta$. The of this  subgroup is then 
given by
\be
\oc=\int d^2a\ d^2 d\ \delta^2(ad-1) =\int \frac{d^2 a}{|a|^2} = 2 \pi \int\frac{dr}{r}\,,
\label{2.12}
\ee
and thus  diverges logarithmically.\footnote{The same   conclusion follows also from the 
definition of $\oc$ as a ratio  $\O_M\ov \O_2$ in \rf{2.8}.}

Let us now  return to the   semiclassical evaluation 
of the 2-point function in  section 3.2 where  \rf{2.1}  and \rf{3.11} (corrected by extra 
``canonical dimension'' $|\xi_1 - \xi_2|^{-4}$ factor)
implies that 
\be
G(\vec{a}_1, \vec{a}_2)\sim \oc^{-1} \frac{1}{|\vec{a}_{12}|^{2\Delta}} \,.
\label{2.18}
\ee
%
This   may seem to  vanish as $\oc$ is divergent. 
However,  we did not yet take into account that 
the  semiclassical solution~\eqref{3.6} is not unique:
it is  defined up to $AdS$   target space $SO(1,5)$  transformations
(acting as  euclidean conformal group at the boundary) 
that preserve the points $\vec{a}_1$, $\vec{a}_2$.\foot{Similar 
 $SO(6)$ degeneracy can be ignored as the  corresponding group has finite volume.}
 This degeneracy requires introduction of the corresponding 
  collective coordinates over which one has to integrate.
  
   Let us count the parameters of these residual symmetry transformations, setting e.g., 
 $a_2=0$ in~\eqref{3.6}. 
First, we have $SO(3)$ rotations in the $(x_1, x_2, x_3)$-plane.
Second,  all translations are broken because they shift the origin and this  
cannot be undone by either boosts or special conformal transformations since they all preserve 
the origin. Now let us act on  $\vec{a}_1=(a_1, 0, 0, 0)$ 
with a dilatation (with parameter $\rho$) and a special conformal 
transformation (with parameters $b_m$):
\be
a_1^{\prime m} =\frac{\rho a_1^m +b^m\rho^2a_1^2}{1+2 \rho b_0 a_1 +\rho^2  b^2 a_1^2}\,.
\label{2.19}
\ee
If all $b_m$ are non-zero the components $a_1^{1}, a_1^2, a_1^3$ will be shifted from zero.
However, they can be moved back to their original values by boosts
in the $(x_0, x_1)$, $(x_0, x_2)$ and $(x_0, x_3)$ planes. 
Thus, we get 4 equations
for 8 parameters ($b_m$, $\rho$, and 3 boosts)
leaving 4 independent parameters. 
Together  with 3 $SO(3)$ rotations this gives 
7 
residual symmetries.  
Note that this number  is just the difference
between the dimension  of $SO(1, 5)$  and the number of the conditions 
set by  fixing  2 points on the boundary, i.e. $15-2 \times 4 =7$. 

Thus  the semiclassical calculation of \ci{bt1} and section 3.2 should include 
the integral over  the corresponding 7 collective coordinates.
The precise form of the integral depends on the location of the two boundary points
but its value  does not, so we may  make a convenient 
choice  of  $\vec{a}_2=0$, $\vec{a}_1=\infty$. 
Then the unbroken subgroup consists
of dilatations and  all $SO(4)$ rotations
(translations are broken because they do not preserve 
the origin and special conformal transformations are broken because they do not preserve infinity). 
Since $SO(4)$ has finite volume, the non-trivial factor comes only from 
the integral over the dilations.
The subgroup of dilatations can be  embedded into 
$SO(1, 5)$ as  diagonal 6-matrices
\be
{\rm diag} ( \rho, \tilde{\rho}, 1,1,1,1)\,,\qquad \qquad \rho \tilde{\rho}=1\,.
\label{2.20}
\ee
The group-invariant  volume  of the corresponding transformations is then 
%
\be
\Omega_{dil} =\int d \rho d \tilde{\rho} {\ } \delta (\rho \tilde{\rho}-1)=
\int \frac{d \rho}{\rho}\,.
\label{2.21}
\ee
This integral is logarithmically divergent  like 
 $\oc$ in~\eqref{2.12}  and thus we may set  $\Omega_{dil} \oc^{-1}=1$
 implying a  finite expression for the 2-point function  in \adss.

The same  argument  applies, in fact, to generic 
 $AdS_{d+1}$  case, e.g., to  strings in $AdS_2\times M$, $AdS_3\times M$ or $AdS_4\times M$.  
  The number of the corresponding collective coordinates is   given  by the 
 dimension of the subgroup of $SO(1, d+1)$ preserving two boundary points  which is 
\be
{\rm dim}[SO(1, d+1)]-2d = \frac{d(d-1)}{2}+1\,.
\label{2.22}
\ee
If we choose the two points to be at $0$ and $\infty$ then the  unbroken subgroup is the 
product of $SO(d)$ and dilatations. The dimensions of these two groups are precisely 
the two terms in the r.h.s. of~\eqref{2.22}.
The integral over the collective coordinates
is again the integral over $SO(d)$ (which gives a finite number)
 times the 1-dimensional integral~\eqref{2.21} over the dilatations.
 It again cancels the diverging  $ \oc$ factor in the 2-point function \rf{2.8}.

Let us 
mention that 
 the divergent integral~\eqref{2.21} may  also be
 interpreted as $\delta (\Delta_2 - \Delta_1) \to  \delta (0)$, 
 like in the Liouville theory~\cite{Seiberg1} and in string theory on $AdS_3$~\cite{ku}.
This argument is not using semiclassical 
approximation 
 and requires a certain analytic continuation.
 Let us  start with the general expression~\eqref{2.1} and single out the integral 
over the dilatations by setting 
 \be z=\rho z^{\prime}\ , \ \ \ \ \ \ \ \ \ x^m =\rho x^{\prime m}
 \ee
  where 
$z^{\prime}$ and $ x^{\prime m}$ are fixed under the dilatations.
 As the string  action~\eqref{2.5} and
$\UU$ in~\eqref{2.3} will  not depend on $\rho$,  we will get
\be  
G \sim \bra ...
\int \frac{d \rho}{\rho} 
\ \Big[\frac{\rho z^{\prime}}{\rho^2z^{\prime 2}+(\rho x^{\prime m}-a^m_1)^2}\Big]^{\Delta_1}
\Big[\frac{\rho z^{\prime}}{\rho^2z^{\prime 2}+(\rho x^{\prime m}-a^m_2)^2}\Big]^{\Delta_2}...\ket 
\,,
\label{n1}
\ee
where $d \rho \ov \rho$ is the group-invariant measure.
To decouple the integral over $\rho$ we may 
again choose the locations of the operators 
at $\vec{a}_1=\infty$ and $\vec{a}_2=0$. Then we will get the factor in \rf{n1}
\be
\hat \Omega_{dil} = \int \frac{d \rho}{\rho}\ \rho^{\Delta_2-\Delta_1}=\int d \eta \ 
e^{(\Delta_2-\Delta_1)\eta}\,, \ \ \ \ \ \ \ \ \ \ \ \   \rho =e^\eta \ . 
\label{n2}
\ee
%
Analytically continuing $\eta \to i \eta$ as in ~\cite{Seiberg1} we
may  interpret this factor as $\delta(\Delta_2-\Delta_1)$, implying that 
the 2-point function   vanishes unless $\Delta_2=\Delta_1$ when 
the singular factor $(\hat \Omega_{dil})_{\Delta_2=\Delta_1}$  gets  cancelled   against
$\oc$ as discussed above.

\

In  the case of the  3-point function \rf{333}
when  3 target space  points $a_i$ are fixed the remaining symmetry 
subgroup of $SO(1,d+1)$ is  compact $SO(d-1)$ and thus  the resulting correlator is finite.
Indeed, let us choose 2 out of 3 fixed 
boundary points to be at $\vec{a}_1=0$  and  $\vec{a}_2=\infty$.
The third point $\vec{a}_3$ breaks 
dilatations and the only surviving symmetry  is the $SO(d-1)$ subgroup of 
$SO(d)$ that  preserves $\vec{a}_3$. 
The same applies of course to higher-point correlators.

\section{Explicit form  of the Schwarz-Christoffel map \\  for  non-extremal  
correlators}

Here we  present  explicit form of the Schwarz-Christoffel transformation found 
by doing the integral in eq.~\eqref{5.7}. 
In our convention the operator with  dimension $\Delta_1 \geq \D_2+ \D_3 $ is inserted at 
$\tau=-\infty$ and the  other two  operators 
 are inserted at $\tau=+\infty$.
We find  (up to an integration constant which we can adjust
 to satisfy \rf{5.11}) 
%
\bea
\tau+ i \s=  \ln {\xi - \xi_1\ov \xi_{12} \xi_{13} M_1}  - \frac{d_2}{d_1} \ln {\xi - \xi_2\ov 
 \xi_{12} \xi_{32}M_2} 
 - \frac{d_3}{d_1} \ln { \xi - \xi_3\ov \xi_{13} \xi_{23}M_3}  \ , 
\ \label{A.1}
\eea
where  \ \ $\xi_{ij} \equiv \xi_i - \xi_j $, 
\bea  
&&M_1=  2 d_1 Q  -  (d_2^2 - d_3^2)  \xi_{23}  (\xi - \xi_1) 
  + d_1^2 \big[  (\xi - \xi_2)\xi_{13} -  (\xi - \xi_3)\xi_{12}  \big]   
  \ ,    \la{a1}\no \\ 
   &&
M_2=  2 d_2 Q  -  (d_3^2 - d_1^2)  \xi_{13}  (\xi - \xi_2) 
  + d_2^2 \big[  (\xi - \xi_3)\xi_{12} -  (\xi - \xi_1)\xi_{32}  \big]   
   \ ,    \la{a2}\no \\ 
  &&
M_3=  2 d_3 Q  -  (d_1^2 - d_2^2)  \xi_{21}  (\xi - \xi_3) 
  + d_3^2 \big[ (\xi - \xi_1)\xi_{32} - (\xi - \xi_2)\xi_{31}   \big]   
    \ ,    \la{a3}  
    \eea 
    and  
\be Q= \Big[d_1^2 \xi_{12}\xi_{13}  (\xi - \xi_2) (\xi - \xi_3) 
 +  d_2^2\xi_{12}\xi_{32} (\xi - \xi_1)(\xi - \xi_3)
  +  d_3^2\xi_{13}\xi_{23} (\xi - \xi_2)(\xi - \xi_1)\Big]^{1/2}
 \,.
\label{A.2}
\ee
The  parameters of the critical point of the map determining 
 the interaction point on
the diagram in Figure 4,  are   determined from a quadratic equation that has two solutions 
\bea
&&
\xi_{int}^{(1)} = \frac{     d_1^2\xi_{12}  \xi_{13} (\xi_{2} + \xi_{3}) + d_2^2\xi_{12}  \xi_{32} (\xi_{1} + \xi_{3})
+ d_3^2\xi_{23}  \xi_{13} (\xi_{2} + \xi_{1}) - P^{1/2} \xi_{23}\xi_{12}      }{2 (d_1^2\xi_{12}\xi_{13}  
 + d_2^2  \xi_{21}\xi_{23}   + d_3^2 \xi_{13}\xi_{23} )} 
 \label{A.3}\\
&&
\xi_{int}^{(2)} = \frac{     d_1^2\xi_{12}  \xi_{13} (\xi_{2} + \xi_{3}) + d_2^2\xi_{12}  \xi_{32} (\xi_{1} + \xi_{3})
+ d_3^2\xi_{23}  \xi_{31} (\xi_{2} + \xi_{1}) + P^{1/2} \xi_{23}\xi_{12}      }{2 (d_1^2\xi_{12}\xi_{13}  
 + d_2^2  \xi_{21}\xi_{23}   + d_3^2 \xi_{13}\xi_{23} )}  \ , 
\label{A.4}\\ 
&& P=  d_1^4 + d_2^4 + d_3^4 - 2 d_1^2 d_2^2 - 2 d_1^2  d_3^2  - 2 d_2^2  d_3^2 
 = - \lambda^{-2}\a_1 \a_2 \a_3 (\a_1+ \a_2+ \a_3) \ , 
\no \la{A5}
\eea
where in the last relation we used that $d_i= {\D_i \ov \sqrt \lambda}$
and the definitions of $\a_i$ in \eqref{1.19}.

\


\end{document}